\documentclass[aps,titlepage,12pt]{revtex4}
\usepackage{amsfonts}
\usepackage{amsmath}
\usepackage{graphicx}
\usepackage{dcolumn}
\usepackage{bm}
\usepackage{overpic}
\usepackage{booktabs}
\usepackage{color}
\usepackage[justification=raggedright,font={small,sf}, singlelinecheck=false]{caption}

\begin{document}
\title{Impactful scientists have higher tendency to involve collaborators in new topics}

\author{An Zeng$^1$, Ying Fan$^1$, Zengru Di$^1$, Yougui Wang$^1$, Shlomo Havlin$^{2,}\footnote{havlin@ophir.ph.biu.ac.il}$}

\affiliation{
$^1$ School of Systems Science, Beijing Normal University, Beijing, China.\\
$^2$ Department of Physics, Bar-Ilan University, Ramat-Gan 52900, Israel.
}

\begin{abstract}
In scientific research, collaboration is one of the most effective ways to take advantage of new ideas, skills, resources, and for performing interdisciplinary research. Although collaboration networks have been intensively studied, the question of how individual scientists choose collaborators to study a new research topic remains almost unexplored. Here, we investigate the statistics and mechanisms of collaborations of individual scientists along their careers, revealing that, in general, collaborators are involved in significantly fewer topics than expected from controlled surrogate. In particular, we find that highly productive scientists tend to have higher fraction of \emph{single-topic} collaborators, while highly cited, i.e., impactful, scientists have higher fraction of \emph{multi-topic} collaborators. We also suggest a plausible mechanism for this distinction. Moreover, we investigate the cases where scientists involve existing collaborators into a new topic. We find that compared to productive scientists, impactful scientists show strong preference of collaboration with high impact scientists on a new topic. Finally, we validate our findings by investigating active scientists in different years and across different disciplines.
\end{abstract}


\maketitle

\section*{Introduction}
Coauthored publications in science have increased significantly during the last century~\cite{the2007wuchty,principles2014milojevic}. Through collaboration, scientists could bring new ideas and techniques from different fields, which in many cases result in high quality publications. Indeed, it has been found that the number of authors in a paper~\cite{large2019wu} as well as the less prior collaboration relations between coauthors~\cite{fresh2021zeng} are strongly associated with the originality of the paper. Thus, scientific collaboration seems to be an important key to enhance innovation of research teams.

Studies regarding scientific collaborations have a long history and have attracted much attention in recent years~\cite{science2018fortunato,the2017zeng}. Early works on scientific collaboration concentrate on collaboration networks constructed from scientific publication data~\cite{the2017zeng}. Numerous topological properties of collaboration networks have been revealed, such as small-world features~\cite{the2001newman}, assortative degree mixing~\cite{assortative2002newman}, rich motifs~\cite{motifs2011krumov}, and community structure~\cite{community2002grivan}. In recent years, attention has been given to further aspects of scientific collaboration. Regarding the collaboration frequency as tie strength, weak, strong, and super strong ties in scientific careers have been identified, and the super ties have been found to have a positive effect on productivity and citations~\cite{quantifying2015petersen}. For coauthored papers, methods have been designed to collectively allocate credits among authors~\cite{collective2014shen}. Another trend to understand collaboration relations is from the perspective of scientific teams, with research questions ranging from team assembly mechanisms~\cite{team2005guimera} to the effect of team characteristics on team performances~\cite{understand2016klug,multinational2015hsiehchen,evolution2016coccia,multi2008jones}. A specific type of collaboration, namely the mentor-mentee relations, has been recently shown to influence research performance~\cite{the2010malmgren} and academic rewards of scientists~\cite{scientific2021jin}.

In recent years, numerous works have been devoted to investigate topic switching in individual careers. With the help of the field classification codes in physics, it has been found that research interest of individual physicists could shift significantly from the beginning to the end of the career~\cite{quantifying2017jia}. The transition map of scientists from field to field has been also extracted from the data~\cite{taking2019battiston}. By applying the community detection technique in the co-citing networks of individual scientist's papers, scientists have been found to have a narrow distribution of the number of major topics during their life time~\cite{increase2019zeng}. This framework has been later used to understand the careers of Nobel laureates~\cite{scientific2020li} and identify the key mechanisms for hot streaks in scientists' careers~\cite{understand2021liu}. However, the characteristics and mechanisms of scientist behind initiating collaboration on a new topic have not been studied. In fact, scientists' choice to collaborate on a new topic is a fundamental process that drives the creativity and impact of the scientific research. The increasingly in-depth development of science requires specialization and accumulated knowledge for researchers to work on a topic~\cite{the2009jones,age2011jones}, suggesting a hypothesis that science might be dominated by single-topic collaboration. On the other hand, inter-disciplinarity and atypical combination of knowledges have been shown to promote creativity~\cite{tradition2015foster,atypical2013uzzi}, suggesting another hypothesis that involving collaborators specialized in a topic to another topic may bring fresh ideas and unexpected solutions. Thus, a series of fundamental questions regarding research topics in scientific collaboration naturally arise: how many different topics do a pair of scientists typically collaborate on? How scientists differ in involving collaborators in their research topics? What factors would affect the probability of a collaborator to join a new topic of a given scientist?

In this paper, we address the above questions by systematically investigating the co-evolution of topics and collaborators during a scientist career, aiming to understand how scientists choose to collaborate on a new topic of research. We decompose the publication series of a scientist to partial series that record the coauthored papers with each of his/her collaborators, allowing us to understand the statistics of the topics that collaborators are involved. The partial time series also enable us to study the temporal features of the collaboration topic formation. By comparing the data of highly productive and highly cited scientists, we investigate how successful scientists of these two types differ in involving their collaborators to new topics. We finally compare active scientists along the past 80 years and across different disciplines, to understand the evolution and disciplinary differences regarding the topics in scientific collaboration.

\section*{Results}
We first describe the method~\cite{increase2019zeng} to identify the involved topics of each collaborator of a focal scientist. The method begins with constructing a network of the focal scientist's publications where the links are defined by the co-citing relations (see Supplementary Fig. S1). We then detect communities in the co-citing network, where each major community represents a different research topic of the focal scientist. In a scientist's publication time series, we mark each paper with a color according to the community it belongs to, see Fig. 1a. The colored time series thus exhibits how a scientist switches from one topic to another. To capture the involved topics of the collaborators, we decompose the series of the focal scientist to partial series, each of which consists of all the coauthored papers with a given collaborator. The topics that a collaborator is involved can be identified by the marked colors of the coauthored papers. In Fig. 1, we illustrate the publication time series of a typical scientist, as well as the decomposed time series of his collaborators. The figure indicates that many collaborators of this scientist are involved in a very small number of his topics.

To statistically test, quantify and understand the pattern illustrated in Fig. 1, we analyzed the scientific publication data of the American Physical Society (APS) journals as well as five other data sets from other disciplines (see details in Methods). The present study will mainly focus on the APS data. The results of the other five data sets are similar to those of APS and are summarized in Fig. 6 and Supplementary Figs. S12-S18. After name disambiguation~\cite{quantifying2016sinatra}, the APS data contains 236,884 distinct scientists. We consider as focal scientists, in order to ensure meaningful community detection results, all scientists that have published at least 50 papers, resulting in 3420 focal scientists. The rest of the authors are included in the analysis, as they may appear as collaborators of these focal scientists.

The first question we ask is, in how many topics the collaborator of a focal scientist is typically involved. To this end, for each focal scientist, we take all his/her collaborators who coauthored at least 2 papers with him/her, and calculate the number of topics that each collaborator is involved. The distribution of collaborators in a number of topics is computed for each focal scientist. Then we evaluate over all focal scientists the average fraction of collaborators for a given number of topics, as shown in the probability distribution in Fig. 2a. The results indicate that on average 63\% collaborators of a scientist are involved in a single topic, and about 25\% are involved in two topics, while 12\% in three and more topics. To test whether this phenomenon can be explained by random behaviour, we consider a surrogate time-controlled reshuffling of the coauthored papers of the collaborators. In the reshuffling process, each paper coauthored by a collaborator and the focal scientist is exchanged with a randomly selected paper that is published in the same year by another collaborator and the focal scientist (see Supplementary Fig. S2 for illustration). By comparing the real data and the controlled surrogate in Fig. 2a, one can see that the high fraction of scientists involved in a single topic, 0.63, cannot be explained by the controlled surrogate, which is significantly smaller, 0.45, suggesting the significant tendency of focal scientists to involve collaborators in fewer topics than expected by the surrogate (for significant test, see Supplementary Fig. S3). To further support this, we calculate for all focal scientists the probability density of the fraction of their collaborators involved in only one topic. As seen in Fig. 2b, the distribution of the fraction of single-topic collaborators follows a roughly normal distribution, with the most probable value around 0.65, very close to the mean value. The surrogate of reshuffled data follows also a roughly normal distribution, yet with a much smaller most probable value, close to 0.4. We further compute the distribution of the number of involved topics for collaborators who coauthored at least 10 papers with the focal scientists. Collaborators with many joint papers have a higher chance to be involved in more than one topic, see Fig. 2c. Nevertheless, despite the smaller fraction of single-topic collaborators in real data, 0.2, it is still much higher, over a factor of 3, than that of reshuffled data, 0.06. We show in Fig. 2d also the distribution of the fraction of single-topic collaborators among those coauthored at least 10 papers with a focal scientist. It can be seen that the distribution is no longer normal as Fig. 2b, and the majority of focal scientists in this case have very low fraction of single-topic collaborators.

The results in Fig. 2a-d also indicate that the number of involved topics is strongly associated with the number of coauthored papers. To quantify this effect, we study directly in Fig. 2e the relation between the number of involved topics and the number of coauthored papers. The results suggest a positive correlation between these two quantities. Note that, the nearly linear relation under logarithmic x-axis indicates that the number of involved topics increases very slowly, i.e., logarithmically, with the number of coauthored papers. Note also, in Fig. 2e, that the number of involved topics in real data is consistently smaller than those of reshuffled surrogate data for different number of coauthored papers. This further supports that collaboration with the same collaborator on several topics is limited i.e., lower than expected in a surrogate control. We also compute, in Fig. 2f, the fraction of single-topic collaborators for collaborators with different number of coauthored papers with the focal scientists. One can see that the fraction of single-topic collaborators decreases with the number of coauthored papers. Nevertheless, the fraction of single-topic collaborators in real data is constantly higher than that in the surrogate data, confirming the tendency of collaborators to join efforts in a single topic.

We further ask how successful scientists are associated with their collaborators in different topics. There are many ways to define a successful scientist. In this paper, we consider two widely adopted metrics, namely the productivity (in terms of total publications) and impact (in terms of the mean citations $c_{10}$ per paper). Here, $c_{10}$ is the number of citations that a paper receives during ten years since it was published~\cite{quantifying2016sinatra}. We show in Fig. 3a that these two metrics are almost uncorrelated, thus selecting the top scientists according to each of the two metrics independently, would result in two very different groups of scientists. Indeed, in Supplementary Fig. S5, we show that the mean citations per paper of the scientists with highest productivity (top 1\%) is roughly the same as the mean citations per paper over all scientists. Also, the productivity of scientists with the highest mean citations per paper (top 1\%) is almost the same as the average productivity of all scientists. In Fig. 3b, we show the relation between the fraction of single-topic collaborators and the number of coauthored papers, and compare between the behaviour of focal scientists with 1\% highest productivity and 1\% highest impact (top 5\% and top 10\% scientists show similar trends, see Supplementary Fig. S6). It is seen that there is no significant difference between these two groups of focal scientists when considering the occasional collaborators (those who coauthored at most 5 papers). However, for the frequent collaborators who coauthored at least 10 papers with the focal scientists (marked by co-pub$\geq10$), it is clearly seen that productive and highly cited scientists behave very differently in involving scientists in research topics. Productive scientists have higher fraction of single-topic collaborators, yet impactful scientists have lower fraction of single-topic collaborators which means higher fraction of multi-topic collaborators. This difference is supported by Fig. 3c where we show directly the distributions of the number of involved topics for frequent collaborators.

To support the finding in Fig. 3b, we calculate the fraction of single-topic collaborators among frequent collaborators (co-pub$\geq10$) for scientists with different productivity and impact in Fig. 3d and 3e, respectively. An increasing trend in Fig. 3d and a decreasing trend in Fig. 3e can be observed. This suggests that the fraction of single-topic collaborators is \emph{positively} correlated with focal scientists' productivity and \emph{negatively} correlated with focal scientists' impact. To quantify the correlation, we directly compute the Kendall's tau correlation ($\tau$) between the fraction of single-topic collaborators (co-pub$\geq10$) of a scientist and the scientist's productivity or impact. The inset of Fig. 3d shows the correlation between the fraction of single-topic frequent collaborators and the focal scientists' productivity, given different impact of the focal scientists. The results suggest that the positive correlation exists even when fixing the impact of the focal scientists, and the correlation is stronger for scientists with smaller impact. The inset of Fig 3e shows the correlation between the fraction of single-topic frequent collaborators and the focal scientists' impact, given different productivity of the focal scientists. The results suggest that the negative correlation exists even when fixing the productivity of the focal scientists, and the correlation is stronger for scientists with higher productivity. In Fig. 3f, we show the fraction of single-topic collaborators (co-pub$\geq10$) of focal scientists with different number of topics. The results indicate that scientists working on more topics tend to have lower fraction of single-topic collaborators. When fixing the number of topics that a scientist has, the fraction of single-topic collaborators of productive scientists is still higher than average, and the fraction of single-topic collaborators of impactful scientists is consistently lower than average.

We further explore the possible reasons leading to the findings in Fig. 3. We first test an interesting hypothesis that our findings are a result of systemic effects that engagement in various fields yields higher impact. If this were the case, the top interdisciplinary scientists would tend to have more citations and higher impact. Their collaborators would, most likely, be interdisciplinary scientists as well, i.e., engaged in multiple topics. However, we find that the mean citations per paper of individual scientists is \emph{negatively} correlated with their number of research topics (see Supplementary Text 4 for more details). This observed pattern suggests that our findings are not due to systemic effects, but more likely a result of individual behavior of scientists. Specifically, a productive scientist is usually a principal investigator and thus has a large research team in which each topic has a specific group of collaborators working on it. This is supported by the evidence in Fig. 4a and 4c that the collaboration network among collaborators of a productive scientist have more significant community structure, which suggests that collaborators of a productive scientist tend to form clusters (possibly according to topics) and they are more likely to work with each other in the same cluster. As the collaborators of a productive scientist tend to work on the topic that they are specialized in, the fraction of single-topic collaborators would be high. On the other hand, the high fraction of multi-topic collaborators of highly cited scientists might be associated with their tendency to work with collaborators who share similar interests. This is indeed supported by the higher fraction of common references between an impactful scientist's papers and his/her collaborators' papers before their collaboration started, as shown in Fig. 4b and 4d. Therefore, the selected collaborators are not only suitable for the initially collaborated topic, but also are preferred collaborators for further topics, which results in higher fraction of multi-topic collaborators.

The next question we ask is what are the features of the collaborators involved in single or multiple topics of a focal scientist. We focus on how the collaboration history with the focal scientist is related to the probability of the collaborator to be involved in the next new topic of the focal scientist. The overall probability of an existing collaborator to be involved in the next new topic of a focal scientist is close to 0.11. We show in Fig. 5a the probability to be involved in the next topic of a focal scientist as a function of the number of past coauthored papers. The results suggest that collaborators who published more papers with a focal scientist have significantly higher probability to be involved in the next topic. Considering that collaborators with many coauthored papers might start collaboration with the focal scientist long ago and may be no longer actively collaborating with the scientist, we further show in Fig. 5a, the probability among recent collaborators who have coauthored papers with the focal scientists within the past two years. The average probability of a recent collaborator appearing in the next new topic of a focal scientist is 0.25, much higher than the overall probability, indicating that a scientist is significantly more likely to involve recent collaborators into a new topic. When considering recent collaborators, the probability to be involved in the next topic still significantly increases with the number of past coauthored papers. The increasing relations can be further quantified by the Kendalls' $\tau$ correlations, given in the legend of Fig. 5a. In Fig. 5c, we also show the correlation for focal scientists with different productivity or impact. One can see that the correlation becomes weaker for productive scientists yet it becomes stronger for impactful scientists. Fig. 5b depicts the relation between the probability to be involved in the next topic and the mean citations of past coauthored papers. Like in Fig. 5a, we compute here also the probability among recent collaborators to become a collaborator of a new topic. Interestingly, in both cases positive correlations are again observed, suggesting that collaborators having published higher impact papers with a focal scientist have significantly higher probability to become involved in the next topic of the scientist. In Fig. 5d, we show that correlation becomes even stronger for impactful scientists. We investigate also the features of selected collaborators for their first topic with a focal scientist, finding similarly that initial collaborators of impactful scientists tend to have much higher mean citations per past paper, up to a factor of 4 compared to low impact scientists (see Supplementary Text 2 and Fig. S8-S9). These results implies that a pair of high impact scientists have significantly higher probability to initiate collaboration on a new topic, compared to a pair of low and high or a pair of low impact scientists.

The observation in Figs. 5c and 5d might be well explainable by the results in Fig. 4. A focal scientist with high productivity usually have a large research team in which each topic has a specific group of collaborators working on it. Therefore, the collaborators are very different in their specialization. When a focal scientist selects collaborators for a new topic, he/she has to take into account both their past performance and their suitability for this topic. Therefore, the productive focal scientists exhibit a lower correlation between the collaborators' past performance and their probability to join the next topic. The impactful focal scientists, on the other hand, tend to work with collaborators who share similar interests to them. The collaborators are generally more likely to be suitable candidates for the new topic of the focal scientists. Taking out the factor of suitability, the past performance of the collaborators thus plays a more important role in affecting the probability to join the next topic. Therefore, for impactful focal scientists, one can observe a higher correlation between the collaborators' past performance and their probability to join the next topic.

Another factor that may affect the probability of collaborators to become involved in a new topic of a focal scientist is the career stage of the focal scientist. We thus show in Fig. 5e the probability of a collaborator to join the next topic in different career stages of the focal scientist. In addition to the overall probability, we provide also the probability of recent collaborators. One can see that the probabilities decrease with the career years of focal scientists, suggesting that scientists in later career stage tend to have lower fraction of multi-topic collaborators (i.e., higher fraction of single-topic collaborators), see Supplementary Fig. S11 for further support. A possible reason for this could be that senior scientists may have research groups each of which consists of specialized collaborators. In Fig. 5f, we compute the Kendall's $\tau$ correlation of the collaborators' past performance (coauthored papers or citations per coauthored paper) and the probability to join next topic in different career stages of the focal scientists. The results show that both correlations decrease with the career years of the focal scientists, suggesting that senior scientists are less sensitive to the past collaboration performance when involving existing collaborators to a new topic.

We further study how the single-topic collaboration phenomenon evolved in the past decades. To this end, we consider focal scientists who started their career in different years and calculate the fraction of their single-topic collaborators. We consider only scientists in their first 30 career years, making scientists who start their career in different years comparable. In Fig. 6a, we observe a decreasing fraction of collaborators involved in a single topic, indicating that in the last century, as time evolved more collaborators of scientists tend to work in multiple topics. Nevertheless, the fraction of scientists involved in a single topic is still significantly higher than surrogate control and the difference seems to be more prominent as time evolved, supporting the significant tendency of single-topic collaborations. We further compare in the inset two groups of scientists whose first 30 career years are respectively from 1940s to 1970s and from 1970s to 2000s. The results show that recent scientists (career from 1970s to 2000s) indeed have a lower fraction of single-topic collaborators for a given number of coauthored papers. In Fig. 6c, we show the average fraction of single-topic collaborators for top-10\% productive and top-10\% impactful scientists whose first 30-year careers are in different periods. One can see that in each time period the impactful scientists have lower fraction of single-topic collaborators compared to overall while the fraction is higher for productive scientists. In Supplementary Fig. S10, we additionally examine the correlation between the probability to join the next topic and the past collaboration performance of a collaborator for focal scientists who started their career in different years. The results show that scientists from different years exhibit similar trends as Fig. 5c and 5d.

Finally, we compare data from different disciplines, including physics, chemistry, biology, computer science, social science and multidisciplinary science. In Fig. 6b, we find a similar form of the distribution of collaborators' topic numbers in different fields. We further find in Supplementary Fig. S12 that the fraction of single-topic collaborators in these disciplines is higher than that of the corresponding surrogate control. The inset of Fig. 6b shows that the fraction of single-topic collaborators is particularly high in biology and chemistry. The reason for this is probably since these two disciplines have many experimentalists whose research fields requires expensive equipments and long-term accumulation of knowledge and master of techniques, which makes them focus on fewer topics (see Supplementary Text 3 and Fig. S13). We additionally show in Fig. 6d that in all considered disciplines impactful scientists have lower fraction of single-topic than overall while productive scientists have higher fraction of single-topic than overall (see Supplementary Fig. S14-S18 for more details).

\section*{Discussion}
Scientific research increasingly depends on teamwork. It is thus critical to understand the collaboration behaviour of scientists. Despite much effort that has been made to investigate the structure and tie strength of collaboration networks, how scientists involve collaborators in their research topics remains poorly understood. In this paper, we find that the actual number of topics in which the same collaborator is involved is significantly smaller than expected from surrogate time-controlled reshuffling, suggesting the preference of recruiting collaborators for a single topic. We interestingly find that productive scientists have higher fraction of single-topic collaborators, yet highly cited scientists have higher fraction of multi-topic collaborators. Our analysis suggest that the observed difference is associated with their tendency in selecting collaborators. The impactful scientists tend to have collaborators sharing similar research interests, while productive scientists tend to have collaborators specialized in a topic. We further study for a focal scientist what are the features of his/her existing collaborators when starting a new topic. We find a stronger tendency of highly cited scientists to involve collaborators with many publications and high citations per paper, yet, in contrast, highly productive scientists have much weaker such tendency. By comparing active scientists in different years, we observe a rising probability, but still significantly smaller than controlled surrogate, of involving collaborators in multiple topics. We finally validate our findings across different disciplines, finding that in all considered disciplines impactful scientists have higher tendency to involve collaborators in new topics.

Our findings can be useful for improving the organization of science. First, our analysis shows that the productivity of a scientist and the average impact per paper of the scientist is almost uncorrelated. Productive scientists usually derive their productiveness from large teams, but our results suggest that these teams do not produce works with above average impact. Therefore, policy makers could consider balancing resources between large and small teams. Secondly, despite that much literature have pointed out that the challenges of the modern world are increasingly interdisciplinary~\cite{the2017zeng}, our work shows that science is still dominated by single-topic collaborations. As multi-topic collaborations are associated with higher impact, proper re-organization of science in terms of encouraging multi-topic collaboration might be helpful for advancing science. Finally, we find that impactful scientists tend to choose impactful scientists as collaborators for a new topic. It implies that successfully breaking new ground is still a task that is hard to be done alone. Thus, there are probably still obstacles to perform interdisciplinary science that need to be removed.

This work may provide a new perspective for understanding individual scientists' careers. In recent years, numerous patterns in individual scientists' careers such as the random-impact rule~\cite{quantifying2016sinatra} and the hot streak~\cite{hot2018liu} have been revealed. However, related analyses inevitably take into account coauthored papers in scientists' career, causing the risk of regarding collaborators' behaviour as the focal scientist's behaviour. It is thus still unclear how to separate the true behaviour of a scientist from the publication records. The method of decomposing publication time series developed in this paper may shed light on this challenging issue. In addition, the framework proposed in this paper can be easily extended to other systems with collaboration such as film actors, patent design, and software development. Finally, we note that our research has limitations. Despite revealing the distribution of topics in scientific collaboration, our work cannot distinguish who is the one initiating their collaboration on these topics. Is it dominated by the focal scientists or by their collaborators? Future investigation on this issue could deepen our understanding on the origin of the observed phenomena in this paper.

\section*{Methods}
\small
\textbf{Data.} We study in this paper six large-scale data sets, including disciplines of physics, chemistry, biology, computer science, social science and multidisciplinary science. The physics data set consists of the scientific publication data of the American Physical Society
(APS) journals~\cite{quantifying2016sinatra}. The computer science data is the AMiner dataset, obtained by extracting scientists' profiles from online Web databases~\cite{extraction2008tang}. The chemistry data contains the publication data of the American Chemical Society (ACS) journals. The biology data contains the publication data of cell publishing group journals. The social science data contains the publication data of SAGE publishing group journals. The multidisciplinary science data contains all papers in five representative multidisciplinary journals including Nature, Science, Proceedings of the National Academy of Sciences (PNAS), Nature Communications and Science Advances. The data of chemistry, biology, social science and multidisciplinary science are extracted according to the DOI of papers from a large publication data set freely downloaded from Microsoft Academic Graph (MAG)~\cite{an2015sinha}. More detailed data description is given in Supplementary Text 1.

\textbf{Decomposing time series.} We first construct for each scientist a co-citing network in which each node is a paper authored by this scientist and two papers have a link if they share at least one reference (see Supplementary Fig. S1). The communities in the co-citing networks are detected via the fast-unfolding algorithm~\cite{fast2008blondel}, with each significant community (more than 5\% papers) representing a major topic of the scientist. As the co-citing network needs to be large enough to ensure meaningful community detection results, we consider only the focal scientists with at least 50 papers. For each focal scientist, we generate the time series presented in Fig. 1a describing the growth history of the network. In the time series, each point is a paper and different colors represent different communities in the co-citing network. Since many of the publications of a focal scientist are resulted from teamwork, the time series is actually aggregated from coauthored papers with different collaborators. We then decompose the publication time series of a scientist to various times series each of which records the coauthored papers with a specific collaborator, as shown in Fig. 1b. The time series of a collaborator clearly exhibits the key information of the collaboration including the number of involved topics, the starting year of the collaboration, the collaboration length, and so on. For better illustration, we show in Fig. 1b the time series of the collaborators with at least 5 coauthored paper with the focal scientist. The illustration of the time series of all collaborators is given in Supplementary Fig. S1. We show also in Supplementary Fig. S4 the statistics of collaboration years and the number of coauthored papers on a topic.

\textbf{Surrogate time-controlled reshuffling.} To examine the significance of an observed pattern in real data, one has to compare it to the result of randomized cases. In this paper, we consider a surrogate time-controlled reshuffling procedure in which the relations between a scientist's collaborators and his/her papers are iteratively randomized. Specifically, a paper coauthored by a collaborator and the focal scientist is exchanged with a randomly selected paper coauthored by another collaborator and the focal scientist. There is time constraint in the procedure that these two papers must be published in the same year, avoiding the case where a collaborator is assigned to a paper that was published even before they started collaboration. In this way, the timing of the collaboration is preserved for each collaborator, yet their involved topics are randomized. The illustration of the surrogate time-controlled reshuffling procedure is presented in Supplementary Fig. S2.

\textbf{Computing the probability to join the next topic.} A scientist may work on multiple topics during his/her career. When a scientist starts to work on a new topic, we calculate the fraction of his/her existing collaborators that will coauthor at least one paper with the scientist in the new topic. The overall probability is obtained by averaging the fraction over all topics except the first topic (as the scientist has no existing collaborators when starting the first topic). One possible concern is that the probability might be underestimated as some collaborators may have already stopped working with the focal scientist long before the focal scientist starts a new topic. We thus further examine the case where all inactive collaborators are removed. Specifically, we calculate the probability to join next topic only among the collaborators who has coauthored at least one paper with the focal scientist in the testing year or one year before.

\textbf{Detecting communities in the collaboration network among collaborators of a scientist.} In Fig. 4, we construct a collaboration network for each focal scientist in which nodes are collaborators of this scientist and links are their coauthorship relations in the scientist's papers. We detect community structure in each of these collaboration networks with the fast unfolding algorithm~\cite{fast2008blondel}. We calculate the maximized modularity $Q_{\rm real}$ of the real networks and the maximized modularity, $Q_{\rm rand}$, in their degree-preserved reshuffled counterparts. The modularity function~\cite{fast2004newman} is defined as
\begin{equation}
Q=\frac{1}{2m}\sum_{i,j}[A_{ij}-\frac{k_ik_j}{2m}]\delta(c_i,c_j),
\end{equation}
where $A$ is the adjacency matrix of the network, $k_i$ is the degree of node $i$, $m$ is the total number of links in the network, $c_i$ is the community to which node $i$ is assigned, the $\delta$ function $\delta(c_i,c_j)$ is 1 for $c_i=c_j$, and 0 otherwise. The communities are obtained when the function $Q$ is maximized.

\clearpage

\noindent  \textbf{Acknowledgments} \\
This work is supported by the National Natural Science Foundation of China under Grant (Nos. 71731002). SH thanks the Israel Science Foundation and the NSF-BSF for financial support.

\noindent \\ \textbf{Author contributions}\\
A.Z. and S.H. designed research; A.Z. performed research; Y.F., Z.D., and Y.W. contributed new reagents/analytic tools; A.Z. and S.H. analyzed data; and A.Z., Y.F., Z.D., Y.W., and S.H. wrote the paper.

\noindent \\ \textbf{Competing financial interests.} The authors declare no competing financial interests.\\

\noindent \textbf{Data and materials availability.} The APS data can be downloaded via \url{https://journals.aps.org/datasets}, the AMiner data can be downloaded via \url{https://www.aminer.cn/aminernetwork}, and the MAG data can be downloaded via \url{https://docs.microsoft.com/en-us/academic-services/graph/}.

\clearpage
\section*{Figures}
\begin{figure*}[h!]
  \centering
  \includegraphics[width=\textwidth]{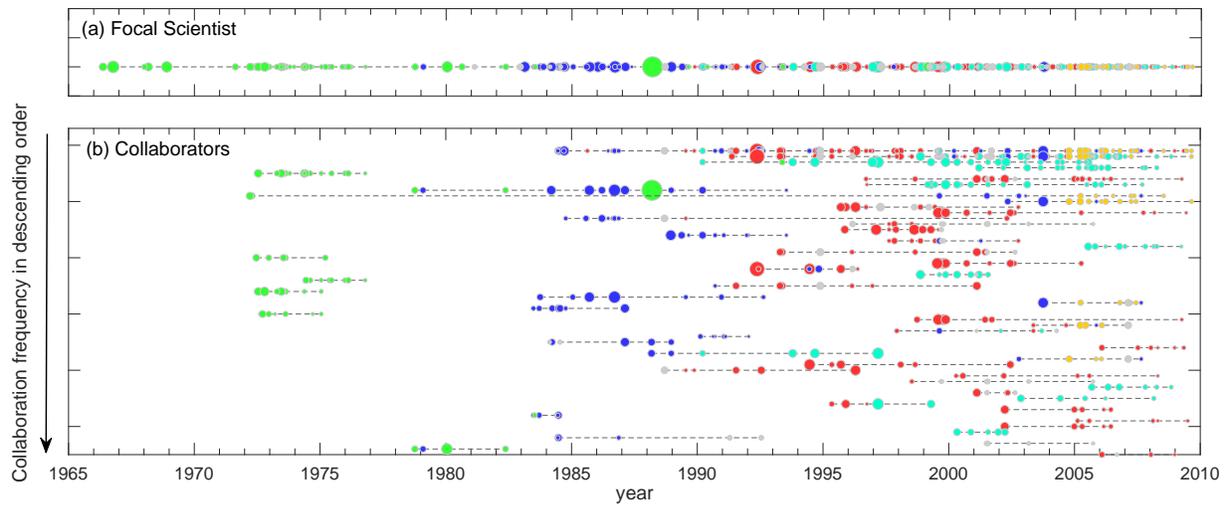}\\
  \caption{\textbf{Illustration of topics that a typical scientist's collaborators are involved.} Panel (a) shows a real typical evolution of research topics during a focal scientist's career. Each node is a paper published by this scientist and the colors of the nodes represent the research topics of these papers. Node size represents the number of citations of this paper. Panel (b) shows the research topics that each of the focal scientist's collaborator is involved. The collaborators are sorted in descending order from top to bottom according to the number of coauthored papers with the focal scientist. Each line shows the results of a collaborator, with each node on it representing a coauthored paper with the focal scientist. Thus the first node and the last node on a line denote respectively the starting year and the ending year of the collaboration. We only show the collaborators who published at least five papers with the focal scientist.}\label{fig1}
\end{figure*}

\clearpage
\begin{figure*}[t!]
  \centering
  \includegraphics[width=\textwidth]{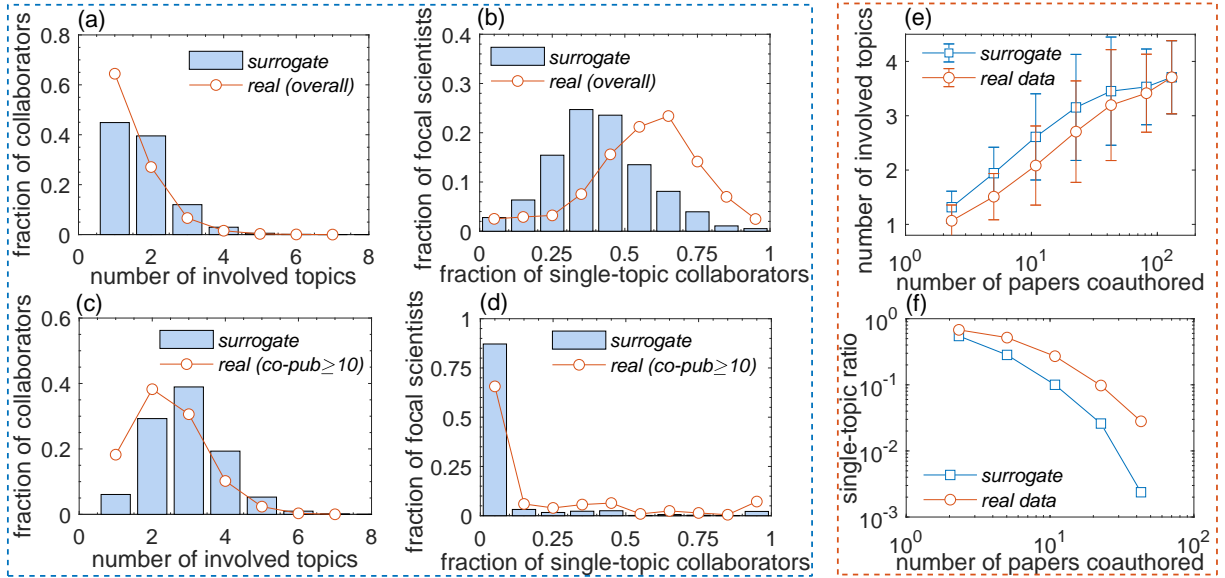}\\
  \caption{\textbf{Number of topics that collaborators are involved.} (a) The distribution of the number of topics that collaborators are involved with a focal scientist. For a scientist, on average about 63\% of his collaborators are involved in only one topic. We also show the results of a controlled surrogate case where the relations between collaborators and their coauthored papers with the focal scientist are randomly shuffled, which is only about 45\%. Note that only the papers published in the same year are allowed to be shuffled in the randomization, see Methods section. (b) For all individual scientists, we calculate the fraction of their collaborators involved in only one topic (denoted as single-topic collaborators). We show in this panel the distribution of the fraction of single-topic collaborators for different scientists. It is clearly seen that the majority of scientists tend to have high fraction of single-topic collaborators compared to the surrogate control. (c) The distribution of the number of involved topics for collaborators who coauthored at least 10 papers with the focal scientists. For a scientist, the fraction of these collaborators involved in one topic is close to 20\%, suggesting that the number of involved topics is strongly associated with the number of coauthored papers. In contrast, the controlled surrogate case has about 6\% single-topic collaborators. (d) The distribution of the single-topic collaborator ratio among those having at least 10 co-publications with a focal scientist. Around 65\% focal scientists have less than 10\% single-topic collaborators. (e) The average number of involved topics for collaborators who coauthored different number of papers with the focal scientist. The result shows that the number of involved topics increases very slowly, i.e., logarithmically, with the number of coauthored papers. The number of involved topics for the surrogate case is higher than that of the real data, again suggesting the strong tendency of scientists to have single-topic collaborators. (f) The fraction of single-topic collaborators for the collaborators who coauthored different number of papers with the focal scientist.}\label{fig2}
\end{figure*}

\clearpage
\begin{center}
  \centering
  \includegraphics[width=16cm]{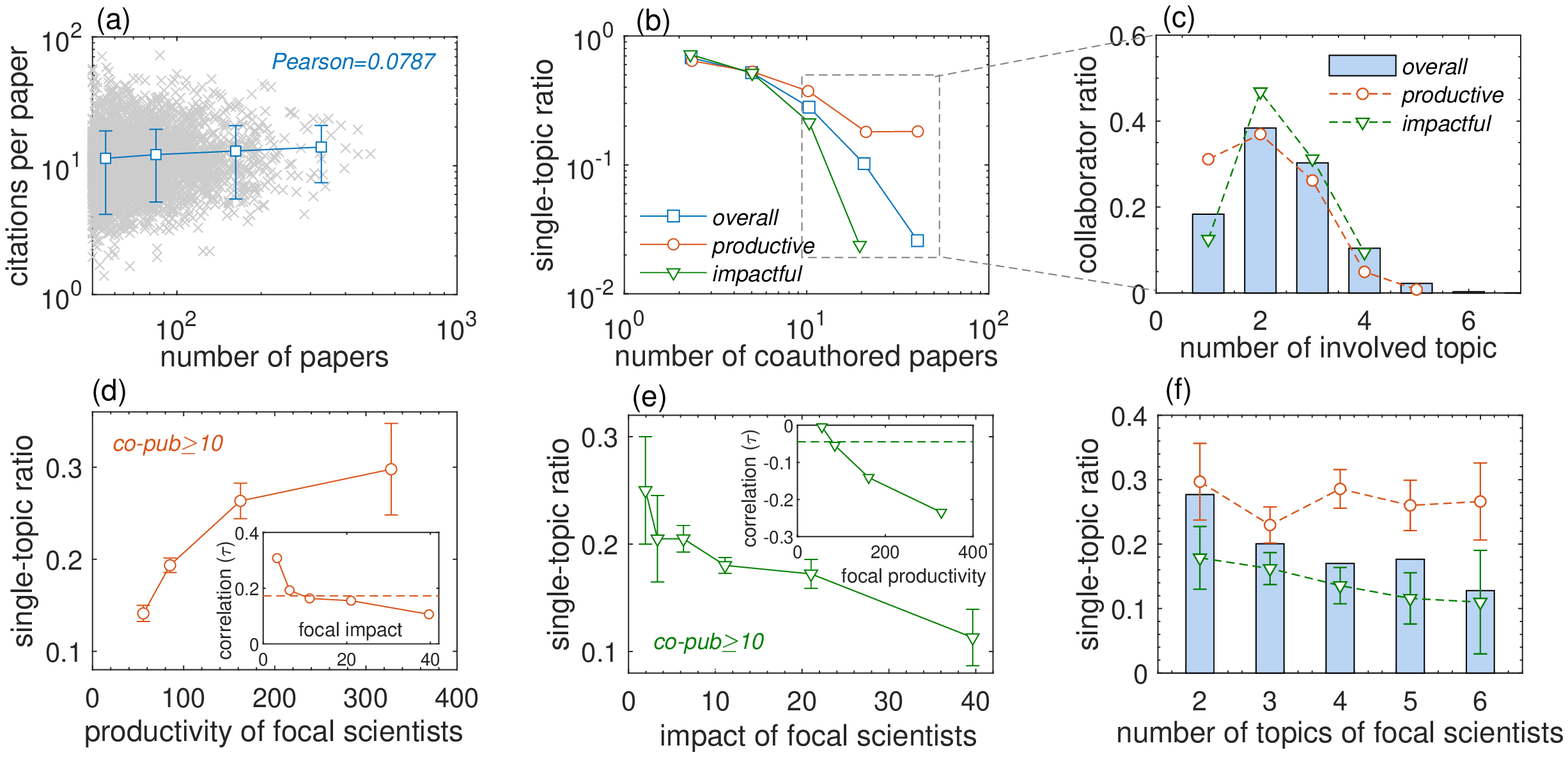}\\
  	\bigskip
	\setbox0\vbox{\makeatletter
		\let\caption@rule\relax
\captionof{figure}[short caption]{\textbf{Productive and impactful scientists associate differently to single-topic collaborators.} (a) Scatter plot of the productivity (measured by the number of papers) and average impact (measured by the mean citations $c_{10}$ per paper) of scientists, where each dot represents a scientist. $c_{10}$ is the number of citations that a paper receives in ten years since it was published. The results show that the correlation between productivity and average impact is very weak, indicated also by the low Pearson correlation of 0.08. Therefore, the scientists with high productivity and the scientists with high impact are two very different groups. (b) The fraction of single-topic collaborators for the collaborators who coauthored different number of papers with the focal scientist. We compare the 1\% most productive scientists (productive in terms of number of published papers) and the 1\% most impactful scientists (impactful in terms of mean citations per paper). (c) The distribution of the number of topics for the collaborators who coauthored at least 10 papers with the focal scientists. The productive scientists have significantly higher fraction of single-topic collaborators while the highly cited scientists have lower fraction of single-topic collaborators. (d) The fraction of single-topic collaborators (among those having at least 10 co-publications) for focal scientists with different productivity. More productive scientists have significantly higher single-topic collaborator ratio. We fix the impact of the focal scientists, and show in the inset the Kendall's $\tau$ \emph{positive} correlation between a scientist's single-topic collaborator ratio and his/her productivity. The dash line in the inset marks the overall correlation. (e) The fraction of single-topic collaborators (among those having at least 10 co-publications) for focal scientists with different impact. Higher impact scientists have significantly smaller single-topic collaborator ratio. We fix the productivity of the focal scientists, and show in the inset the Kendall's $\tau$ \emph{negative} correlation between a scientist's single-topic collaborator ratio and his/her impact. The dash line in the inset marks the overall correlation. (f) The fraction of single-topic collaborators (among those having at least 10 co-publications) for focal scientists with different number of topics. The legend in panel (f) is the same as that in panel (c).}
\global\skip1\lastskip\unskip
		\global\setbox1\lastbox
	}
	\unvbox0
	\setbox0\hbox{\unhbox1\unskip\unskip\unpenalty
		\global\setbox1\lastbox}
	\unvbox1
	\vskip\skip1
\end{center}\label{fig3}

\clearpage
\begin{figure}[t!]
 \centering
  \includegraphics[width=12cm]{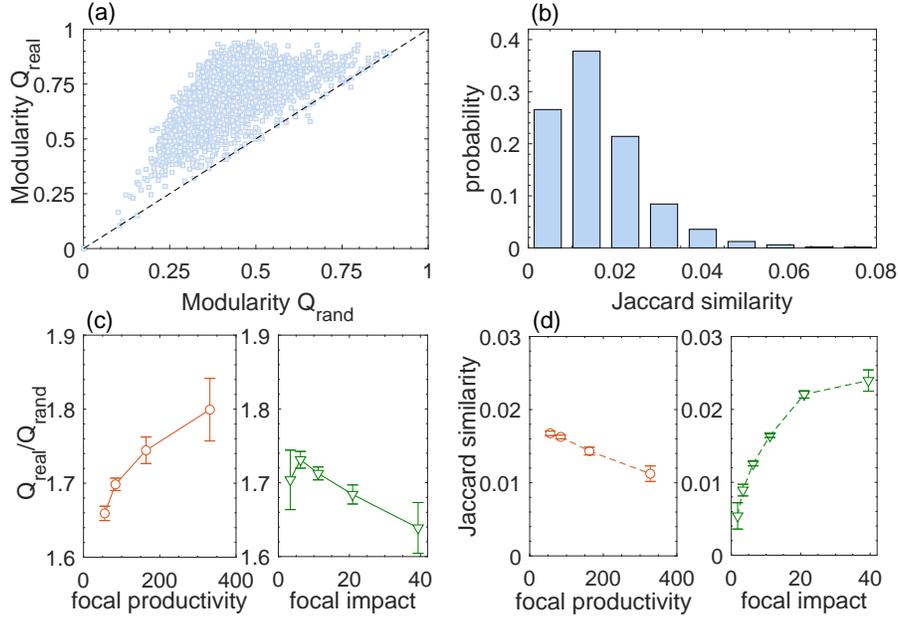}\\
  \caption{\textbf{Features of the collaborators of productive scientists and impactful scientists.} (a) The relation between the maximized modularity $Q_{\rm real}$ and $Q_{\rm rand}$ for all focal scientists. Each pair of $Q_{\rm real}$ and $Q_{\rm rand}$ are obtained by detecting community structure in the collaboration network among collaborators of a scientist and in the degree-preserved reshuffled counterparts, respectively (see Methods). All the points are located above the diagonal line $Q_{\rm real}=Q_{\rm rand}$, indicating that the community structure in real networks is truly significant. (c) $Q_{\rm real}/Q_{\rm rand}$ for focal scientists with different productivity or impact. A larger $Q_{\rm real}/Q_{\rm rand}$ indicates a more significant community structure. (b) To quantify the research interest similarity between a focal scientist and each of his/her collaborators, we measure the Jaccard similarity of the references given by their papers before collaboration (see Supplementary Fig. S7 for results of other similarity metrics). We show the distribution of the mean similarity for all focal scientists. (d) The mean reference similarity for focal scientists with different productivity or impact. Productive scientists and their collaborators have limited research interest in common while impactful scientists and his/her collaborators have more common research interest.}
\end{figure}

\clearpage
\begin{figure*}[t!]
  \centering
  \includegraphics[width=\textwidth]{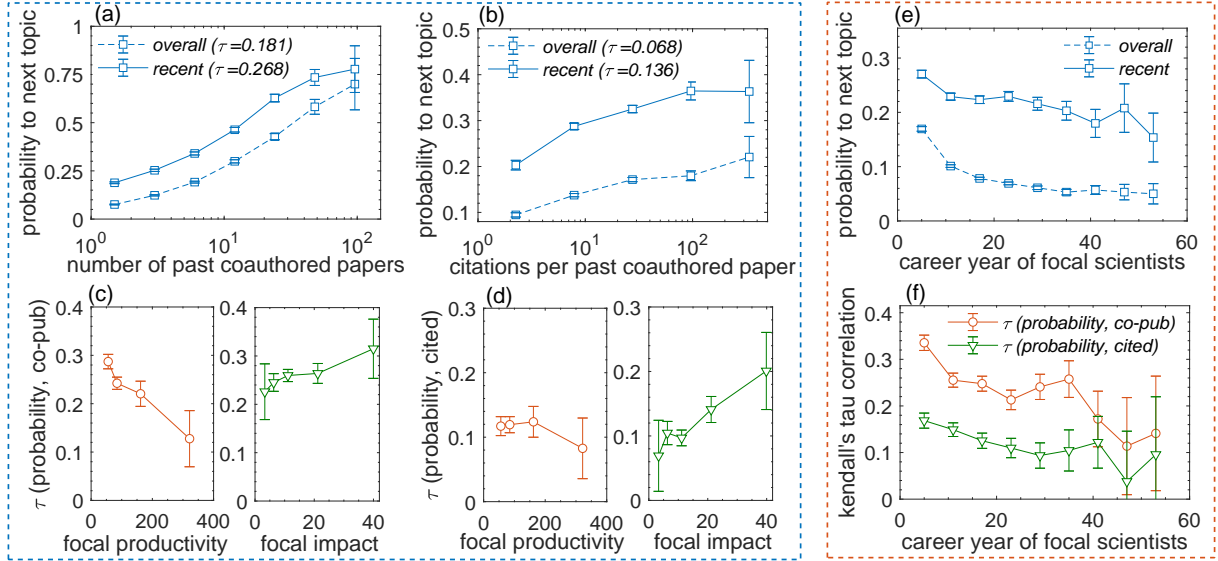}\\
   \caption{\textbf{Factors associated with the probability of an existing collaborator to join a new topic of a focal scientist.} (a) For a focal scientist that starts to work on a new topic, we calculate the probability of his/her existing collaborators to join him in the new topic. The overall probability of an existing collaborator to join the new topic of a focal scientist is close to 0.11. We compute also the probability among recent collaborators who have coauthored papers with the focal scientists within the past two years, and find it to be much higher (i.e. the mean is close to 0.25). Both probabilities show an increasing trend with the number of past coauthored papers, indicating that more intensive past collaboration increases the probability of a collaborator to join a new topic of a focal scientist. (b) The probability of a collaborator to join the next topic of the focal scientist versus the mean citations of their past coauthored papers. Both the overall probability and the probability among recent collaborators (which is much higher) show an increasing trend, suggesting that the collaborators who published highly cited papers with the focal scientist have higher probability to join next topic of the focal scientist. (c) The Kendall's $\tau$ correlation between the probability to join next topic and the number of past coauthored papers of a collaborator, for focal scientists with different productivity or impact. (d) The Kendall's $\tau$ correlation between the probability to join next topic and the citations per past coauthored paper of a collaborator, for focal scientists with different productivity or impact. (e) The probability of a collaborator to join the next topic in different career stages of the focal scientists. Both the overall probability and the probability among recent collaborators show a decreasing trend, suggesting that the focal scientists tend to have higher fraction of single-topic collaborators in their later careers. (f) The Kendall's $\tau$ correlation of the collaborators' performance and the probability to join next topic in different career stages of the focal scientists. The correlations tend to be weaker in the later career stage of the focal scientists.}\label{fig5}
\end{figure*}

\clearpage
\begin{center}
  \centering
  \includegraphics[width=16cm]{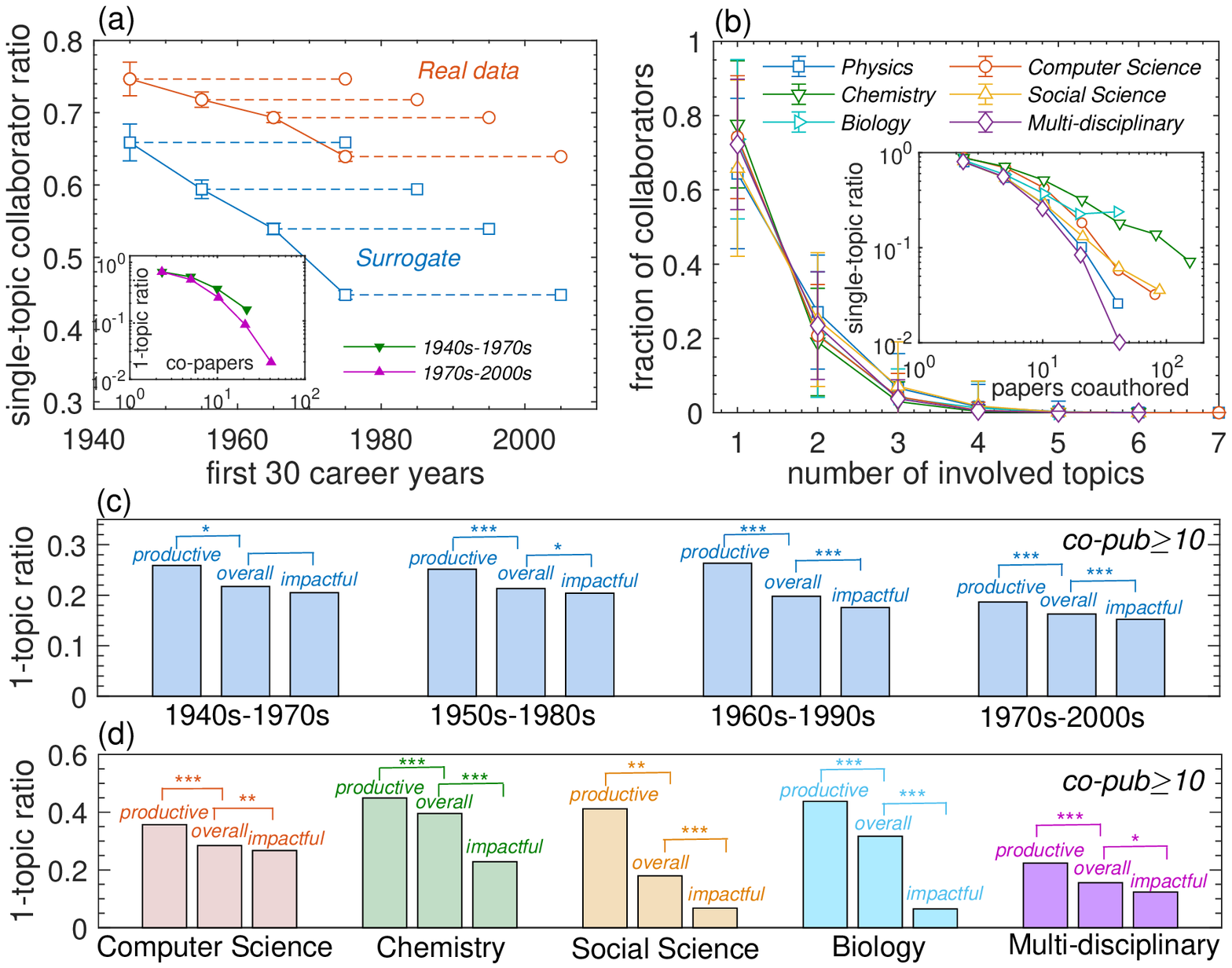}\\
  	\bigskip
	\setbox0\vbox{\makeatletter
		\let\caption@rule\relax
\captionof{figure}[short caption]{\textbf{Evolution in the last century and discipline comparison.} (a) The fraction of single-topic collaborators of scientists who started their career in different years. We consider only scientists' first 30 career years, making scientists that started their careers in different years comparable. The career starting years are marked by symbols and the first 30 career years are denoted by the dash lines. One can see that in more recent years scientists have lower fraction of single-topic collaborators, yet the fraction is always significantly higher than surrogate control. The observed trend is supported by the inset where we show the fraction of single-topic collaborators for the collaborators who coauthored different number of papers with the focal scientist. We compare two groups of scientists whose first 30 career years are respectively from 1940s to 1970s and from 1970s to 2000s. (b) The distribution of the number of topics that collaborators are being involved. We compare data from different disciplines, including physics, chemistry, biology, computer science, social science and multidisciplinary science. Inset: the fraction of single-topic collaborators for the collaborators who coauthored different number of papers with the focal scientist. We observe a strong tendency of single-topic collaboration, with the fraction of single-topic collaborators being particularly high in biology and chemistry (see inset). (c) The average fraction of single-topic collaborators for top 10\% productive and top 10\% impactful scientists whose first 30-year careers are in different periods. (d) The average fraction of single-topic collaborators for productive scientists and impactful scientists in different disciplines. In (c) and (d), the fraction of single-topic collaborators is calculated among collaborators who coauthored at least 10 papers with the focal scientists. Asterisks between two adjacent bars indicate the $p$ values from the Kolmogorov-Smirnov test of the corresponding distributions. Here, $*$, $**$ and $***$ stand for $p\leq0.1$, $p\leq0.01$ and $p\leq0.001$, respectively. Almost all pairs of distributions significantly differ from one another. The large $p$ values for $1940s$-$1970s$ in (c) are due to the small sample sizes.}
\global\skip1\lastskip\unskip
		\global\setbox1\lastbox
	}
	\unvbox0
	\setbox0\hbox{\unhbox1\unskip\unskip\unpenalty
		\global\setbox1\lastbox}
	\unvbox1
	\vskip\skip1
\end{center}\label{fig6}

\clearpage
\begin{center}
{\large\bfseries Supplementary Information}\\[8pt]
{\large Impactful scientists have higher tendency to involve collaborators in new topics}\\[8pt]
\small An Zeng, Ying Fan, Zengru Di, Yougui Wang, Shlomo Havlin\\
\end{center}

\section*{Supplementary Note 1. Data description.}
We study in this paper six large-scale data sets, including disciplines of physics, chemistry, biology, computer science, social science and multidisciplinary science. The physics data set consists of the scientific publication data of the American Physical Society
(APS) journals, with 236,884 authors and 482,566 papers, ranging from year 1893 to year 2010. In the APS data, there are 3,420 authors with at least 50 papers. The computer science data is obtained by extracting scientists' profiles from online Web databases. It contains 1,712,433 authors and 2,092,356 paper, ranging from year 1948 to year 2014. There are 9,818 authors in this data with at least 50 papers. The chemistry data contains the publication data of the American Chemical Society (ACS) journals, with 1,918,866 authors and 1,320,333 papers, ranging from year 1879 to 2020. There are 7,555 authors in this data with at least 50 papers. The biology data contains the publication data of cell publishing group journals, with 432,880 authors and 154,233 papers, ranging from year 2003 to year 2020. There are 218 authors in this data with at least 50 papers. The social science data contains the publication data of SAGE publishing group journals, with 1,909,545 authors and 1,354,511 papers, ranging from year 1965 to year 2020. There are 772 authors in this data with at least 50 papers. The multidisciplinary science data contains all papers in five representative multidisciplinary journals including Nature, Science, Proceedings of the National Academy of Sciences (PNAS), Nature Communications and Science Advances. The dataset consists of 1,077,399 authors and 633,808 papers, ranging from year 1869 to year 2020. There are 1,209 authors in this data with at least 50 papers. The data of chemistry, biology, social science and multidisciplinary science are extracted according to the DOI of papers from a large publication data set freely downloaded from Microsoft Academic Graph.

\section*{Supplementary Note 2. The initial topic that collaborators work with a focal scientist.}
We investigate respectively what are the features of selected collaborators for their first topic with a focal scientist. We study in Supplementary Fig. S8 the past research performance of collaborators before collaborating in their first topic with a scientist. In Supplementary Fig. S8a, we show the distribution of collaborators' past career years when they started collaborating with a focal scientist. The overall distribution has a large value for zero past career, suggesting that almost 40\% collaborators of a focal scientist are new comers to research, i.e., graduate students. The rest of the distribution follows approximately an exponential decay with their past career years. The productive and impactful scientists exhibit similar distributions as the overall. The robustness of this finding is supported by Fig. S8d where we directly calculate the mean past career years of collaborators when joining scientists of different productivity (left panel) or impact (right panel) for new collaborations. In both sub-panels, no significant trend is observed from close to 8 years, indicating that productive and impactful scientists do not show different preference of collaborators at different career stages when selecting new collaborators.

In Supplementary Fig. S8b, we show the distribution of the number of collaborators' past publications when they started working with the focal scientist. The overall distribution exhibits a long-tail, with the distributions of the productive and impactful scientists overlapping with the overall distribution. We also compute in Supplementary Fig. S8e the mean past publications of collaborators when focal scientists with different productivity or impact include new collaborators. The mean past publications of collaborators also seems to be not associated with focal scientists' productivity and only increases weakly with focal scientists' impact.

We additionally show in Supplementary Fig. S8c the distribution of citations per past paper of collaborators just before they started working with the focal scientist. The distribution of impactful scientists exhibits significant difference (green-shifted towards higher impact) from those of the overall and productive scientists. Interestingly, it is seen that high impact focal scientist can be associated with recruiting initial collaborators of higher impact. To further support this feature, we calculate in Supplementary Fig. S8f the mean citations per past paper of collaborators when scientists with different productivity or impact select new collaborators. The focal scientists' productivity is again not associated with the collaborators' mean citations per past paper, yet, interestingly, initial collaborators of impactful scientists tend to have much higher mean citations per past paper, up to a factor of 4 compared to low impact scientists. Thus, our results suggest that a pair of high impact scientists have significantly higher probability to initiate collaboration, compared to the pair of low impact scientists. In Supplementary Fig. S9, we show that the feature of impactful scientists choosing impactful scientists as collaborators is consistent over the years.

\section*{Supplementary Note 3. Relation between research topics and disciplines}
In general, only some disciplines have significant advantage to reach out to other disciplines, eventually resulting in interdisciplinary research. However, in our analysis we actually investigate research topics that are smaller than research fields and disciplines. For instance, within the field of statistical physics, a scientist could study the topics of self-organized criticality, percolation, Ising model and so on. Therefore, even if working in a specific research field or discipline a scientist could have multiple topics in his/her research. We support this argument by detecting topics that individual scientists studied within different disciplines, finding consistently that a scientist could have multiple topics (see Supplementary Fig. S13). Meanwhile, we observe the heterogeneity in topic number across disciplines. Scientists working in methodological science, such as physics, tend to have more topics. Scientists working in experimental science, such as chemistry and biology, tend to have fewer topics.

Apart from the data of individual disciplines, we also investigate in Supplementary Fig. S13 a multidisciplinary data which contains papers in five representative multidisciplinary journals including Nature, Science, Proceedings of the National Academy of Sciences (PNAS), Nature Communications, and Science Advances. We detect topics that individual scientists studied in these multidisciplinary journals, obtaining again multiple topics for different scientists. In this data, topics might be from different disciplines, and multi-topic scientists in this case could be regarded as interdisciplinary scientists.

Our findings are general. In both the disciplinary data and the multidisciplinary data, we observe a consistent pattern that impactful scientists have higher tendency to involve collaborators in new topics, which has been discussed and shown in the manuscript (see Fig. 6).

\section*{Supplementary Note 4. The findings are not a result from systemic effects.}
One possible concern is that our findings may be a result of systemic effects rather than of individual behavior. It could be that the engagement in various fields, which creates additional citations and, hence, additional overall impact. If this were the case, the top interdisciplinary scientists would tend to have more citations and higher impact. Their collaborators would, most likely, be interdisciplinary scientists as well, i.e. engaged in multiple topics.

We thus investigate whether engagement in various research fields would lead to higher impact. We directly calculate the Pearson correlation coefficient of the mean citation ($c_{10}$) per paper of individual scientists and their number of research topics obtained by clustering each individual's papers according to common references (see Methods and Supplementary Fig. S1). The correlation coefficient in the APS data is -0.171 and the $p$-value is smaller than 0.001, suggesting a significant negative correlation. To further support the result, we calculate also the correlation coefficient in the multidisciplinary data, obtaining the correlation coefficient as -0.0627 and the $p$-value as 0.0015. Similar phenomenon has also been observed in a very recent paper which finds that the impact of the new research steeply declines the further the scientists move from their prior work~\cite{adapt2022hill}.

The consistent negative correlations above suggest that engagement in more research fields is not associated with higher impact. This is probably because attention (citations) is connected with reputation in a field~\cite{reputation2013petersen}, and it is harder to establish reputation if the research effort of a scientist is distributed in various research fields. As engagement in more research fields is not associated with higher impact, we believe that the observed patterns in our paper is not due to systemic effects. Instead, it is more likely a result of individual behavior of scientists.

\clearpage
\section*{Supplementary Figures}
\begin{figure}[h!]
  \centering
  \includegraphics[width=\textwidth]{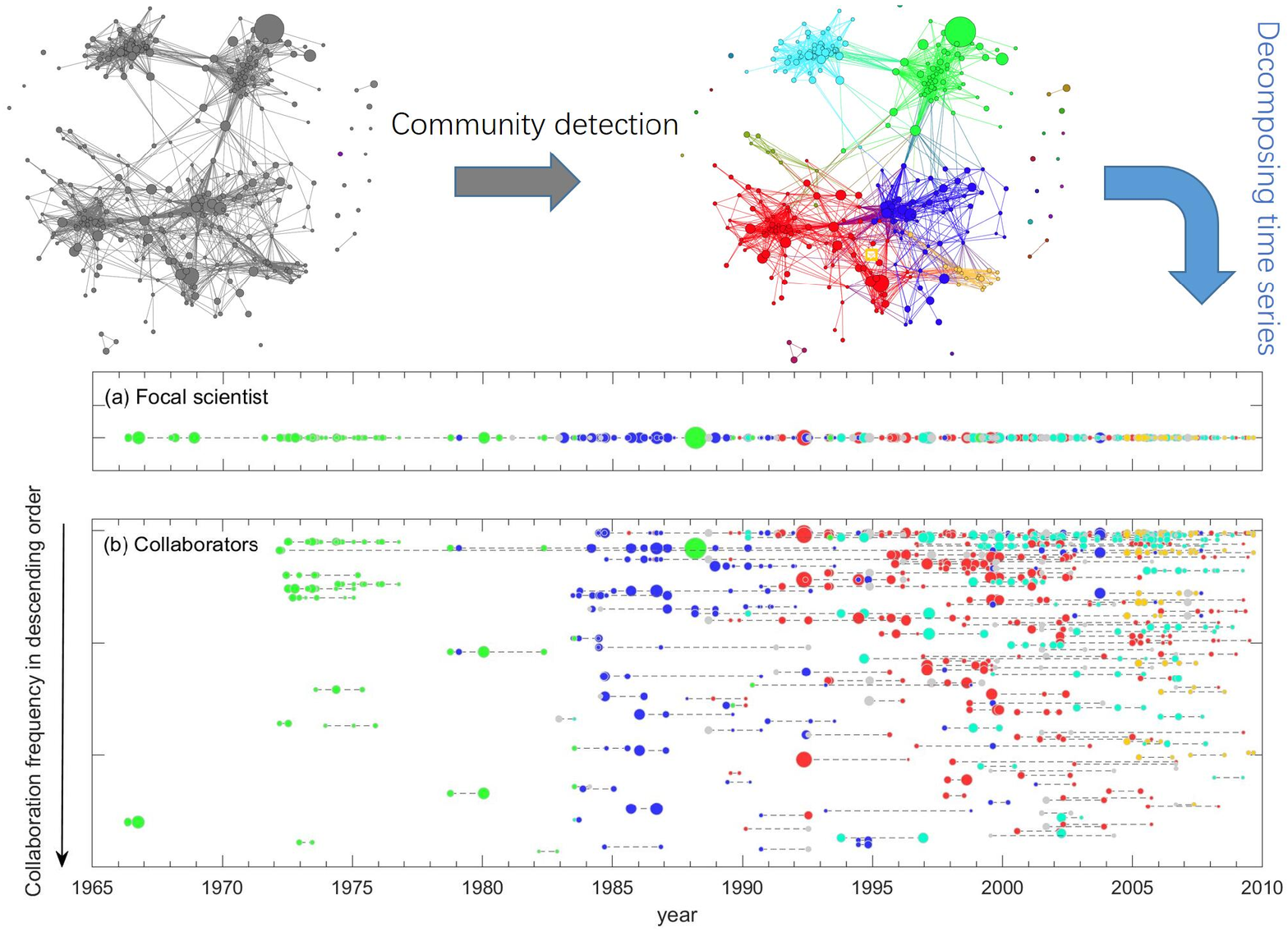}
  \textbf{Figure S1. Illustration of topic detection.} For each focal scientist, we construct a co-citing network in which each node represents a paper of the scientist and two papers are connected if they share at least one reference. This figure illustrate the co-citing network of a typical highly cited scientist. The communities in the network are then identified via the fast-unfolding algorithm, with each community consisting of more than 5\% nodes regarded as a major topic of the focal scientist. The detected communities are then used to color the publication time series of the scientist. Each paper is given to the color according to the community it belongs to, as shown in Fig. S1a. We then decompose the colored time series to various ones with each one containing the coauthored papers with each individual collaborator. The time series of the collaborators clearly demonstrate the number of topics that each collaborator is involved in.
\end{figure}

\clearpage
\begin{figure}[h!]
  \centering
  \includegraphics[width=15cm]{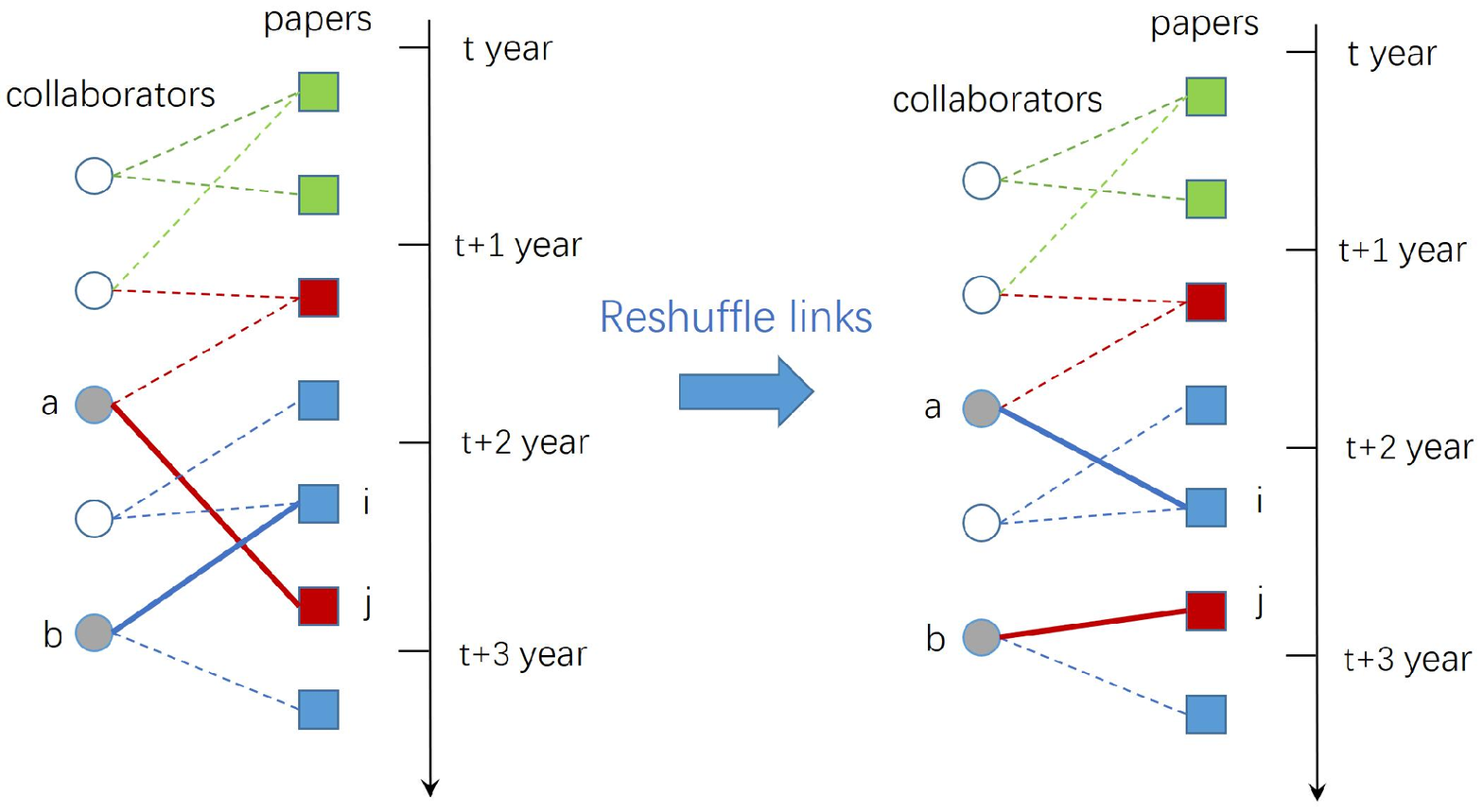}\\
  \textbf{Figure S2. Illustration of time-controlled reshuffling.} In order to examine the significance of the observed patterns in collaboration topics, we compare the real data with the randomized counterparts obtained through time-controlled reshuffling. We first construct a bipartite network for each focal scientist, in which one type of nodes are collaborators of the scientist and the other type of nodes are their coauthored papers. The links in the bipartite network represent the authorship relations between collaborators and papers. To reshuffle the network, we randomly select two links a-j and b-i (paper i and paper j must be published in the same year) and change the connections as a-i and b-j. The reshuffling procedure is repeated for 4 times of the total number of links to ensure that the network is fully randomized. The time-controlled reshuffling preserves the number of coauthored papers of each collaborator, the number of authors in each paper, the total number of topics of the focal scientist, as well as the collaboration timing between each collaborator and the focal scientist. However, the number of involved topics of each collaborator might be altered. By comparing the real data with the reshuffled data, one can see whether the empirical patterns can be explained from the controlled random.
\end{figure}

\clearpage
\begin{figure}[h!]
  \centering
  \includegraphics[width=\textwidth]{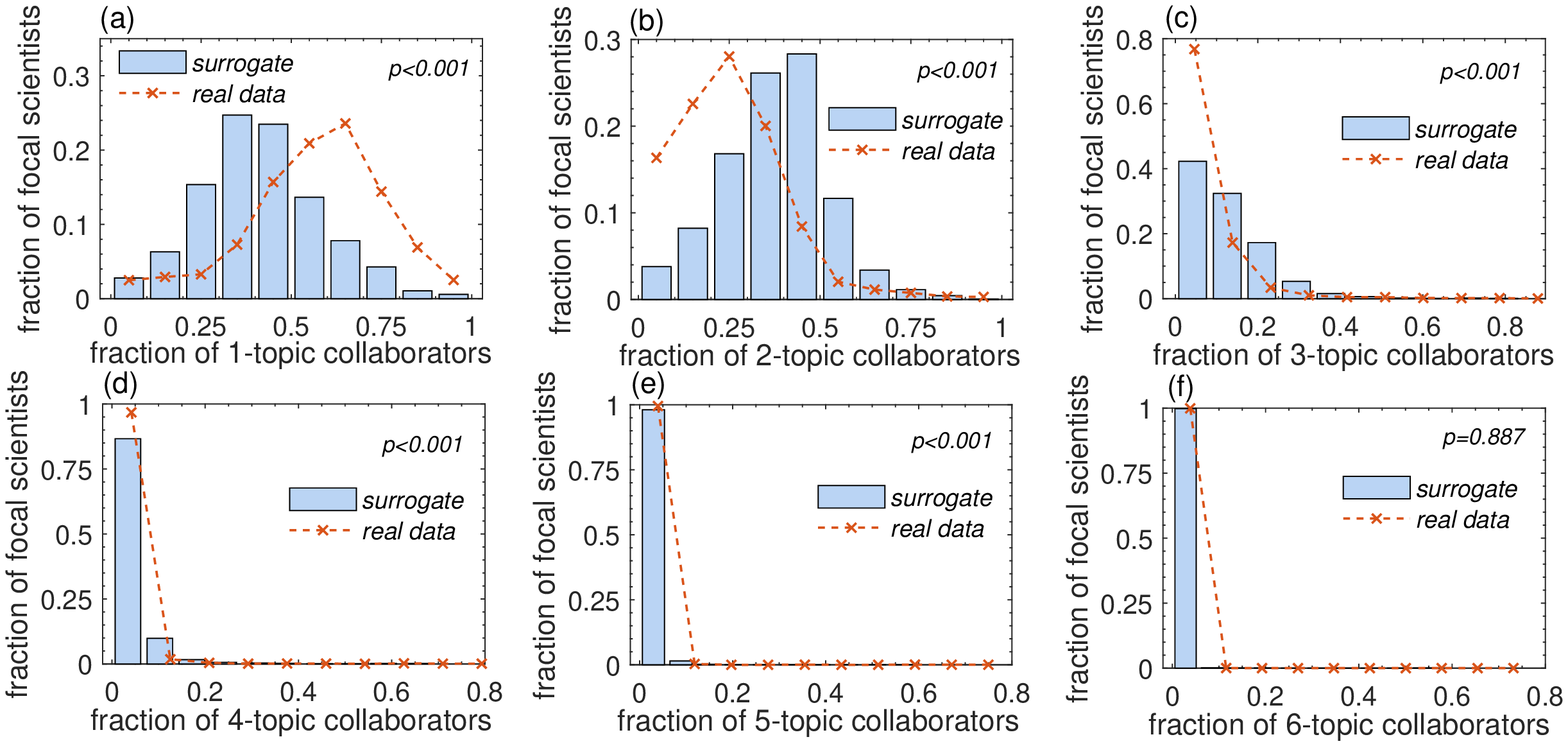}\\
  \textbf{Figure S3. Significant test.} We quantify the significance of the difference between collaborators' involved topic number in real data and time-controlled reshuffled data. (a) The distributions of the fraction of one-topic collaborators in real data and surrogate time-controlled reshuffled data. The distribution describes the fraction of focal scientists having a specific fraction of one-topic collaborators. One can see that the real data and surrogate reshuffled data exhibit a large difference, with the significance supported by the small $p$ value ($p<0.001$) of the Kolmogorov-Smirnov test distinguishing these two distributions. (b-f) The distributions of the fraction of two-topic, three-topic, four-topic, five-topic and six-topic collaborators in real data and surrogate time-controlled reshuffled data, respectively. The $p$ values of the Kolmogorov-Smirnov test distinguishing the distributions of real data and the surrogate reshuffled data in (b-e) are all smaller than 0.001, indicating the significant difference. However, the $p$ value in (f) is 0.887, suggesting no significant difference. This is because the collaborators involved in 6 topics are very few in both real data and surrogate reshuffled data.
\end{figure}

\clearpage
\begin{figure}[h!]
  \centering
  \includegraphics[width=\textwidth]{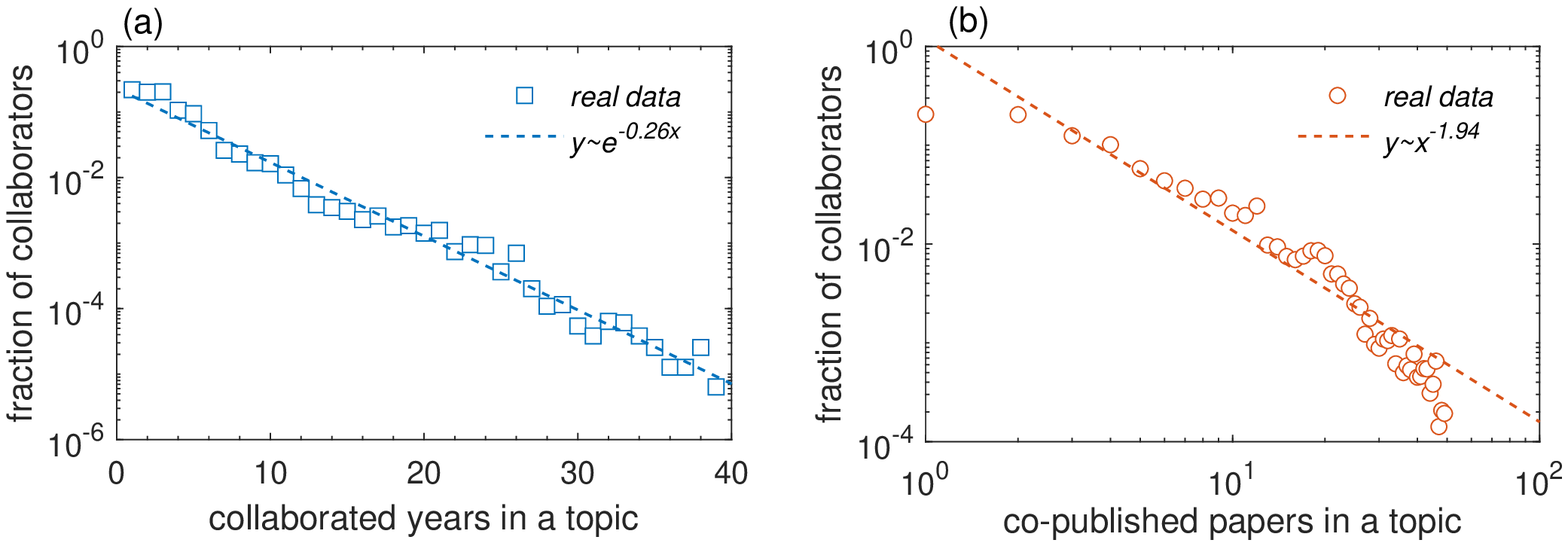}\\
  \textbf{Figure S4. Topic collaboration time and topic co-published papers.} (a) The distribution of collaboration years on a topic. The distribution exhibits an exponential form, indicating a typical collaboration length on a topic (3.85 years). (b) The distribution of the number of coauthored papers on a topic. The distribution exhibits a power-law form, suggesting absent of typical number of coauthored papers on a topic.
\end{figure}

\clearpage
\begin{figure}[h!]
  \centering
  \includegraphics[width=\textwidth]{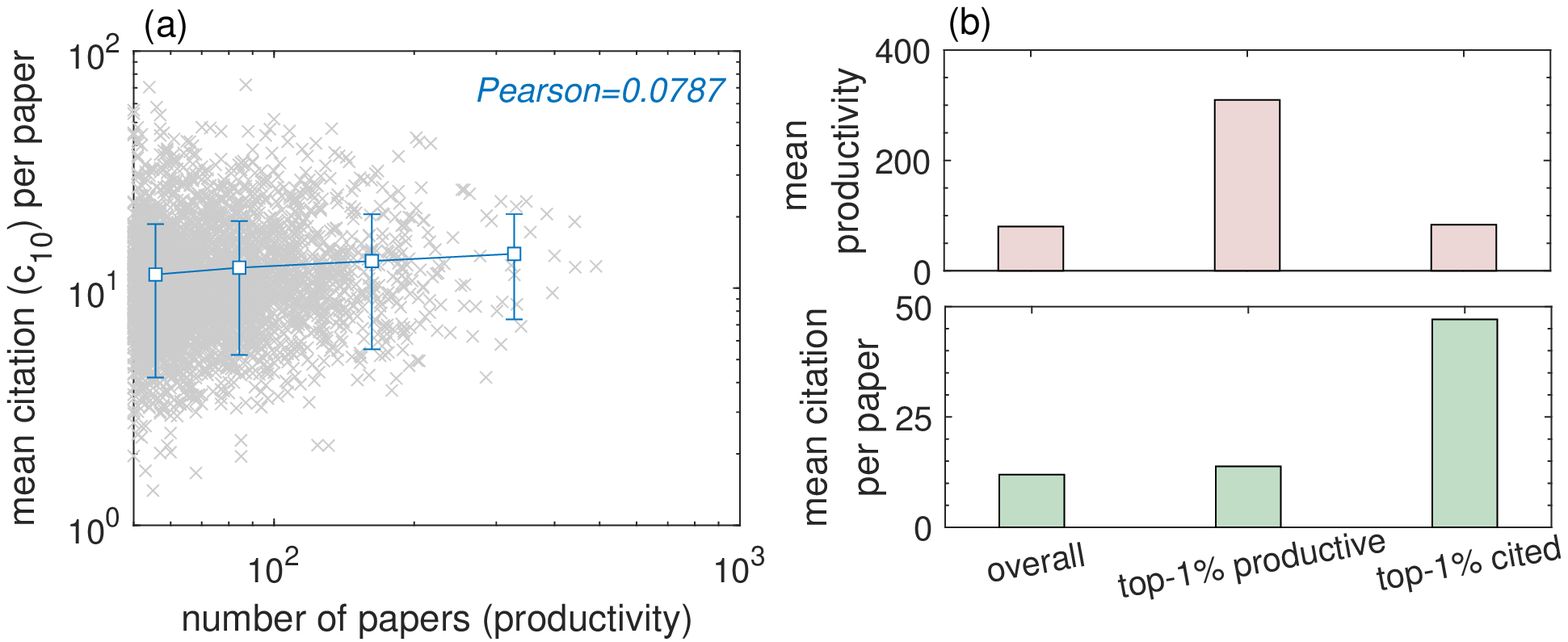}\\
  \textbf{Figure S5. Relation between productivity and mean citation per paper.} (a) Scatter plot of the productivity (measured by the number of papers) and average impact (measured by the mean citation $c_{10}$ per paper. Each point is the results of a scientist. $c_{10}$ is the number of citations that a paper receives until it is published for ten years. The results show that the productivity and average impact are almost uncorrelated, indicated also by the low Pearson correlation (0.0787). Therefore, the scientists with high productivity and the scientists with high average impact are two very different groups of scientists. (b) upper panel: the mean productivity of all scientists (overall), the mean productivity of the top-1\% productive scientists, and the mean productive of the top-1\% average impact scientists. One can see that the mean productivity of the top-1\% average impact scientists is close to the overall mean productivity. Bottom panel: the mean citation per paper of all scientists (overall), the mean citation per paper of the top-1\% productive scientists, and the mean citation per paper of the top-1\% average impact scientists. One can see that the mean citation per paper of the top-1\% productive scientists is close to the overall mean citation per paper.
\end{figure}

\clearpage
\begin{figure}[h!]
  \centering
  \includegraphics[width=\textwidth]{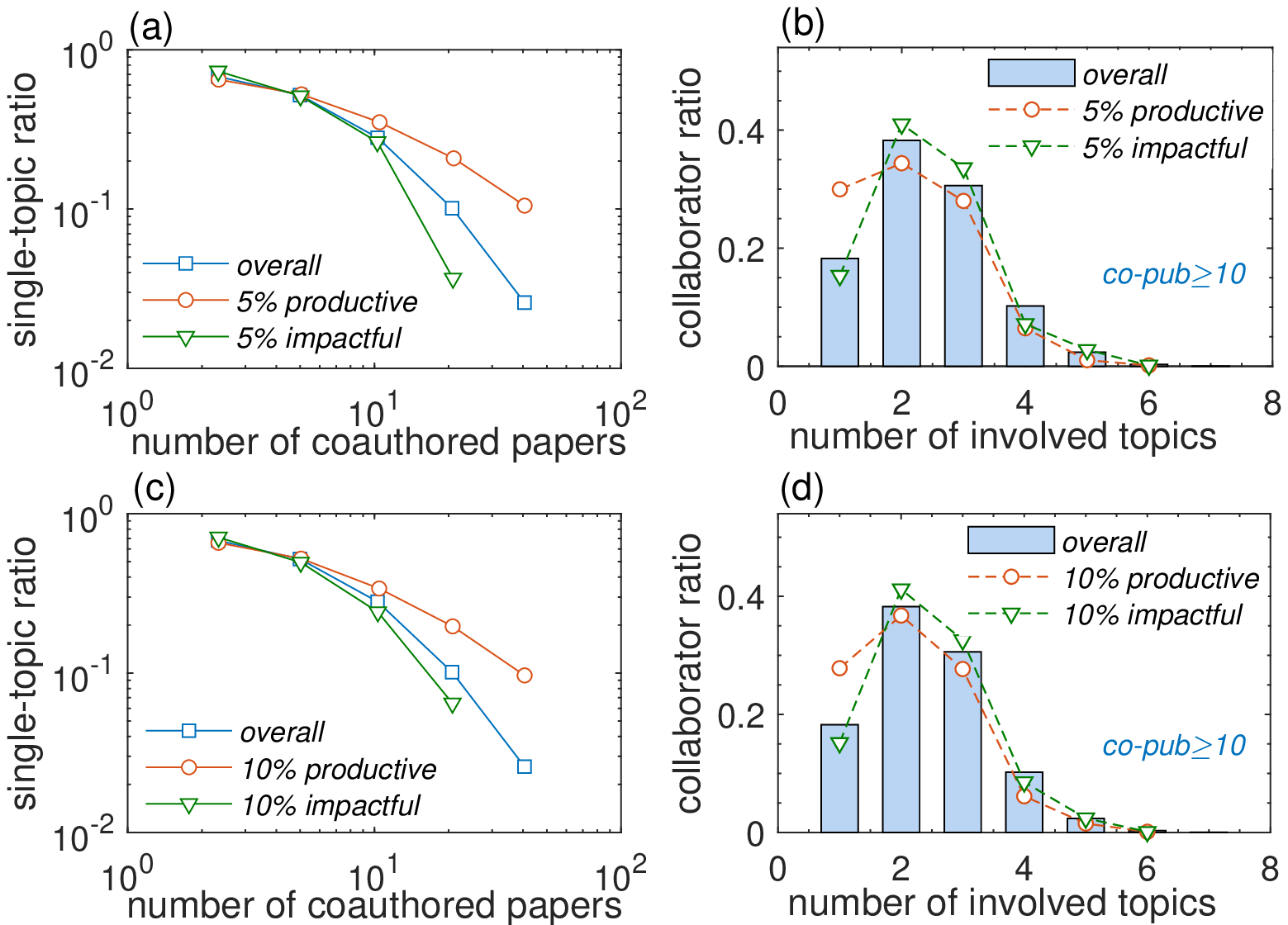}\\
  \textbf{Figure S6. top-5\% and top-10\% scientists with high productivity or high impact.} (a)(c) The fraction of single-topic collaborators for the collaborators who coauthored different number of papers with the focal scientist. (b)(d) The distribution of the number of topics for the collaborators who coauthored at least 10 papers with the focal scientists. In (a)(b), we compare the overall with the 5\% most productive scientists (productive in terms of number of published papers) and the 5\% most impactful scientists (impactful in terms of mean citation per paper). In (c)(d), we compare the overall with the 10\% most productive scientists and the 10\% most impactful scientists. In both cases, the productive scientists have higher fraction of single-topic collaborators while the highly cited scientists have lower fraction of single-topic collaborators.
\end{figure}

\clearpage
\begin{figure}[h!]
  \centering
  \includegraphics[width=\textwidth]{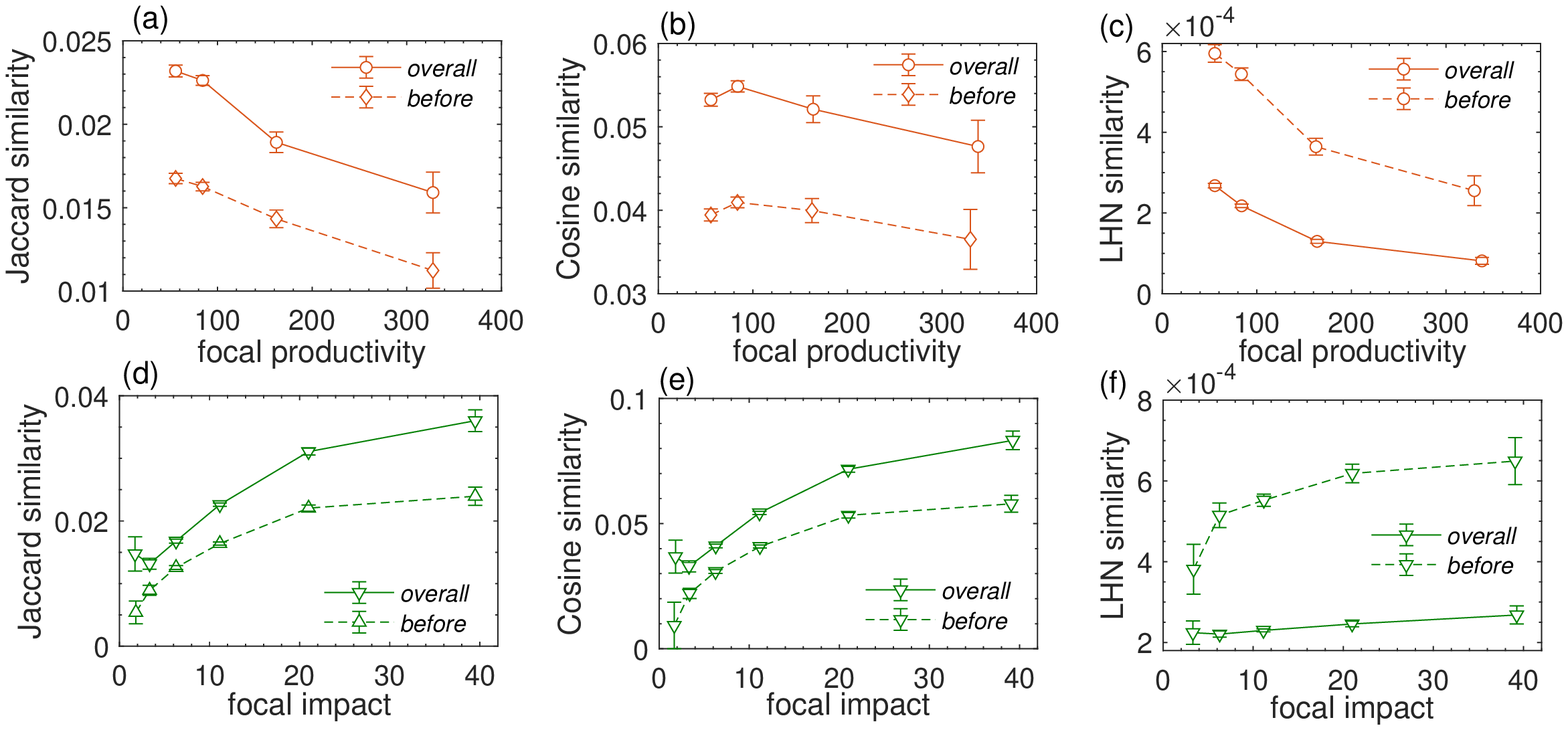}\\
   \textbf{Figure S7. Common references between a focal scientist's papers and his/her collaborators' papers.} We show in Fig. 3 that impactful scientists have higher fraction of multi-topic collaborators. The high fraction of multi-topic collaborators of highly cited scientists might be associated with the tendency to work with collaborators who share similar interests with them. To quantify the research interest similarity between two scientists, we measure the similarity of the references given by their papers. Specifically, we calculate the reference similarity between a focal scientist and each of his/her collaborators, and then compute the average. The ``overall" reference similarity is computed by taking all the papers of two scientists but excluding their coauthored papers. The ``before" reference similarity is computed by taking all the papers of two scientists before they started collaboration. We consider three different similarity metrics, including Jaccard in panel (a)(d), Cosine in panel (b)(e) and Leicht-Holme-Newman (LHN) in panel (c)(f). The Jaccard similarity is $|\Gamma_i\cap\Gamma_j|/|\Gamma_i\cup\Gamma_j|$ where $\Gamma_i$ represents the set of references given by the considered papers of $i$. The Cosine similarity is $|\Gamma_i\cap\Gamma_j|/\sqrt{|\Gamma_i|\times|\Gamma_j|}$. The LHN similarity is $|\Gamma_i\cap\Gamma_j|/(|\Gamma_i|\times|\Gamma_j|)$. (a)-(c) show the relation between the reference similarity and focal scientists' productivity. A decreasing trend suggests that productive scientists and their collaborators have limited research interest in common. (d)-(f) show the relation between the reference similarity and focal scientists' impact. An increasing trend suggests high fraction of common research interest between an impactful scientist and his/her collaborators.
\end{figure}

\clearpage
\begin{figure*}[h!]
  \centering
  \includegraphics[width=\textwidth]{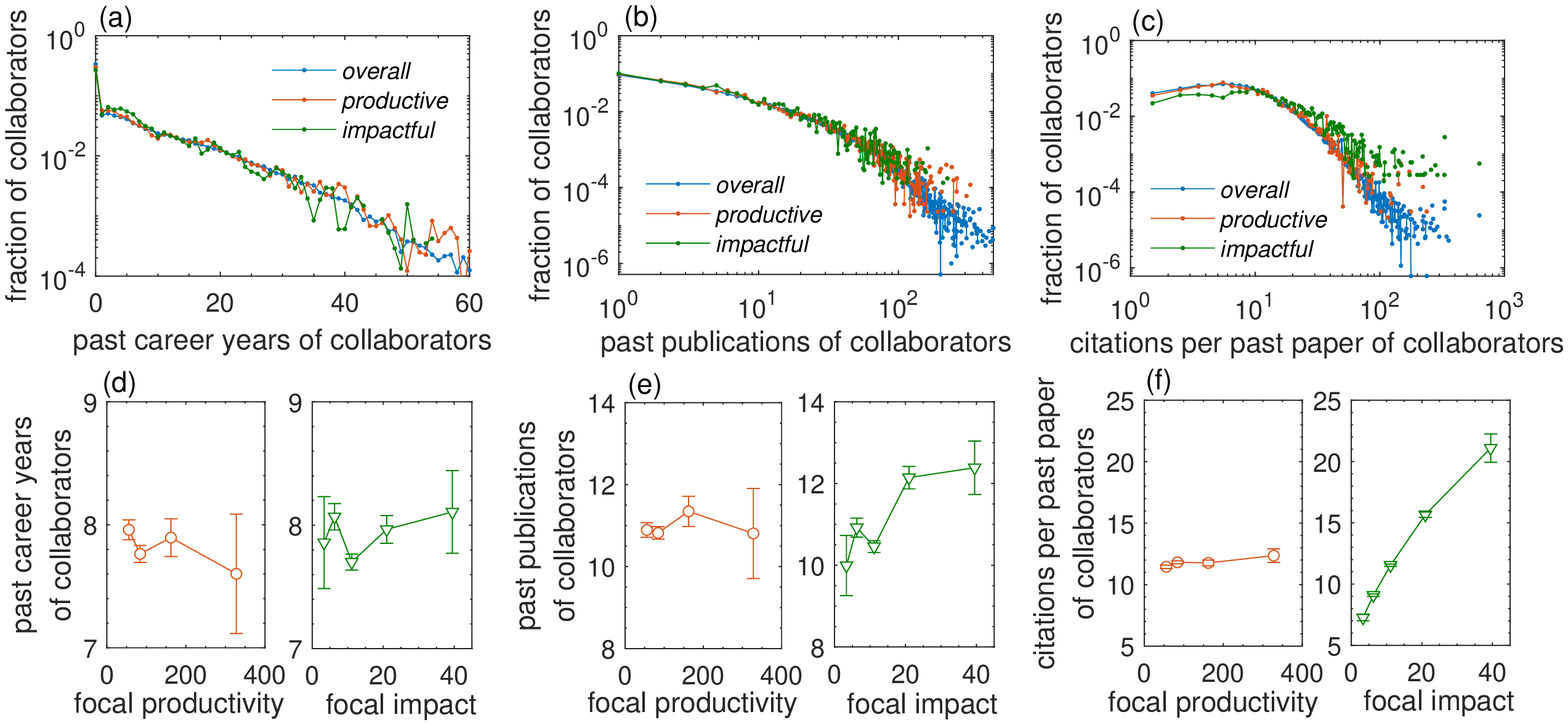}\\
   \textbf{Figure S8. Factors affecting the probability of collaborators to join their first topic with a focal scientist.} (a) For each collaborator, we record his number of career years before he started collaboration with the focal scientist. This panel shows the distribution of the number of past career years of the collaborators, before starting to collaborate with the focal scientist. The distribution shows an exponential tail, with almost 40\% of collaborators are new comers, i.e., mainly graduate students. Note that both the productive and impactful scientists exhibit similar distributions as the overall. (b) The distribution of past publications of collaborators when starting the collaboration with the focal scientist. No significant difference can be observed between the overall and the successful scientists (i.e. productive and impactful). (c) The distribution of citations per past paper of recruited new collaborators just before starting collaboration with the focal scientist. The distribution of productive scientists overlaps with the overall distribution. However, the distribution of impactful scientists exhibits a fatter tail (see Supplementary Table S1 for significance test). This means that impactful scientists are associated with impactful initial collaborators. (d) The average past career years of the collaborators for focal scientists with different productivity or impact. (e) The average past publications of collaborators for focal scientists with different productivity or impact. (f) The average citations per past paper of collaborators for focal scientists with different productivity or impact. In (d)(e)(f), only the average citations per past paper of collaborators have an significant increasing trend with the impact of the focal scientists. This further supports that a high impact scientist can be associated with high impact collaborators and that a pair of high impact scientists have higher probability to initiate collaboration, compared to the pair of low and high or low impact scientists.
\end{figure*}

\clearpage
\begin{figure*}[h!]
  \centering
  \includegraphics[width=\textwidth]{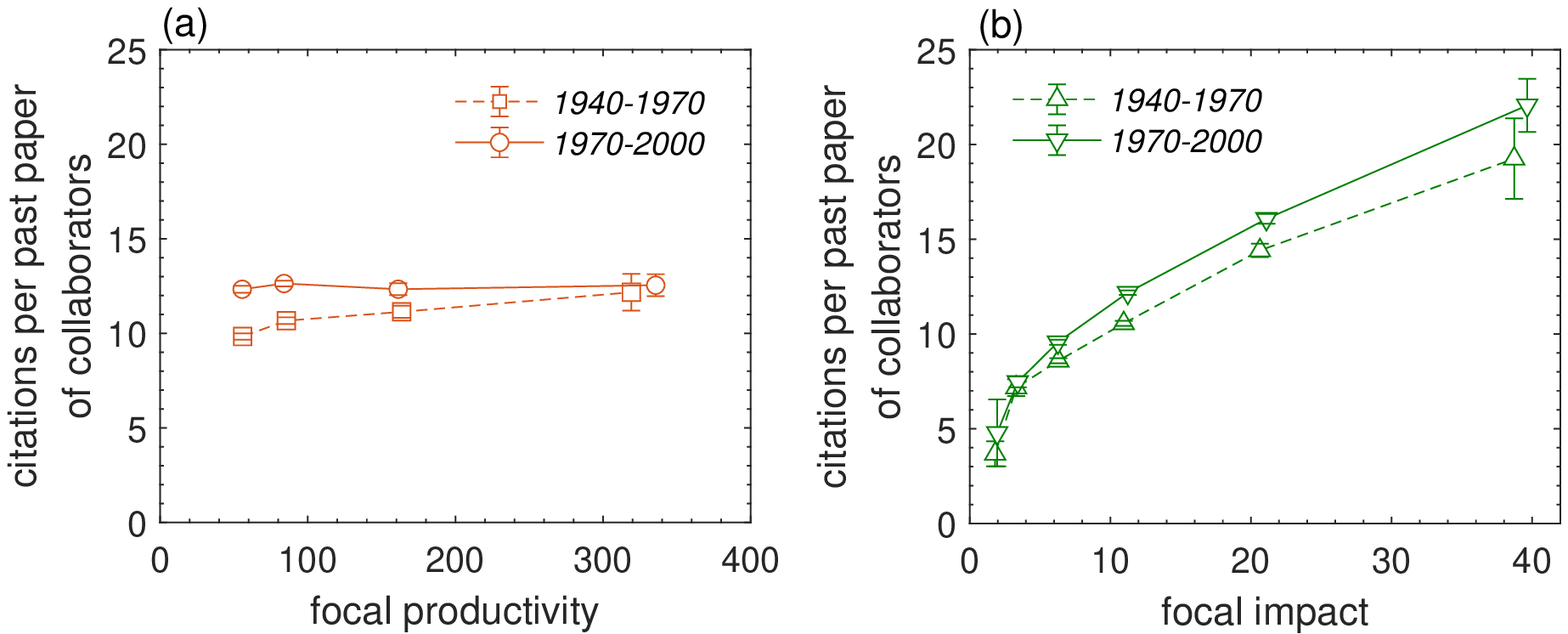}\\
  \textbf{Figure S9. Feature of impactful scientists choosing impactful scientists as collaborators.} (a) The average citations per past paper of collaborators for focal scientists with different productivity. (b) The average citations per past paper of collaborators for focal scientists with different impact. We compare focal scientists who started their career in different years (i.e. 1940-1970 and 1970-2000). In both group, the average citations per past paper of collaborators have an significant increasing trend with the impact of the focal scientists. The results suggest that the feature of impactful scientists choosing impactful scientists as collaborators is consistent over the years.
\end{figure*}

\clearpage
\begin{figure*}[h!]
  \centering
  \includegraphics[width=\textwidth]{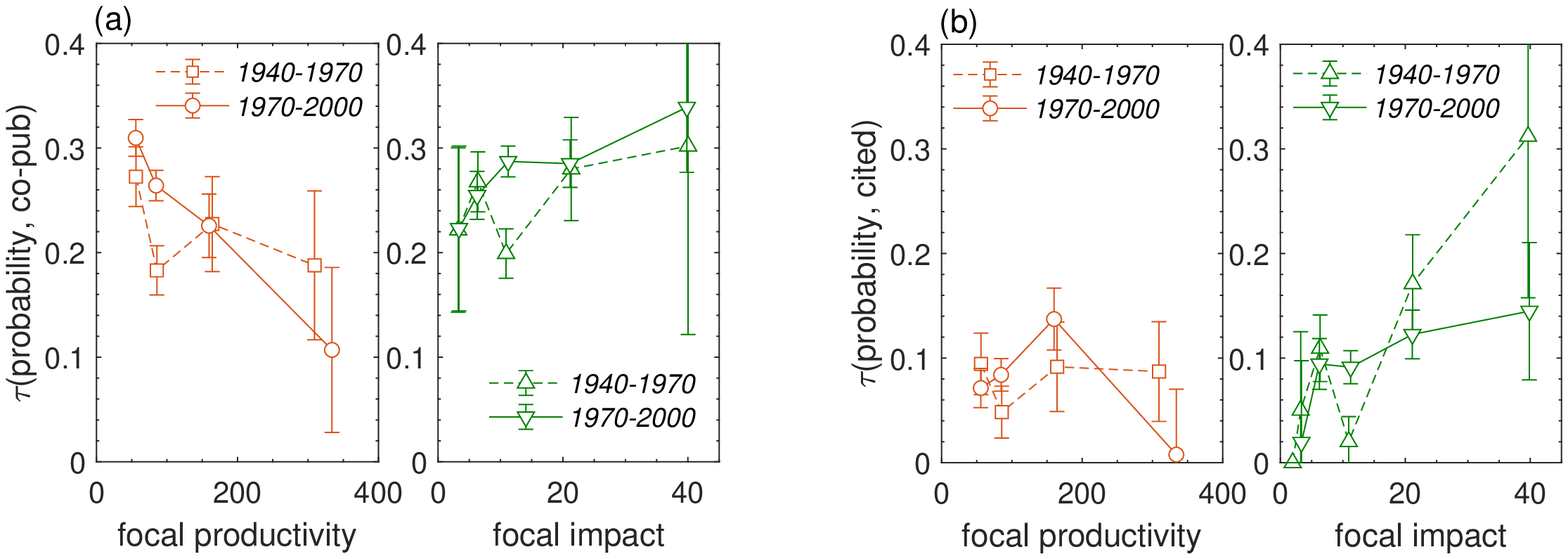}\\
 \textbf{Figure S10. Difference between productive scientists and impactful scientists in different years.} (a) The Kendall's $\tau$ correlation between the probability to join next topic and the number of past coauthored papers of a collaborator, for focal scientists with different productivity or impact. (b) The Kendall's $\tau$ correlation of the probability to join next topic and the citations per past coauthored paper of a collaborator, for focal scientists with different productivity or impact. Here, we compare the scientists who started their career between 1940 and 1970, and the scientists who started their career between 1970 and 2000. In general, the productive focal scientists exhibit a lower correlation between the collaborators' past performance and their probability to join the next topic. For impactful focal scientists, one can observe a higher correlation between the collaborators' past performance and their probability to join the next topic. One can see that the two groups of scientists exhibit similar trends. The larger fluctuation in the early group (1940-1970) is due to the smaller sample size in this group.
\end{figure*}

\clearpage
\begin{figure}[h!]
  \centering
  \includegraphics[width=\textwidth]{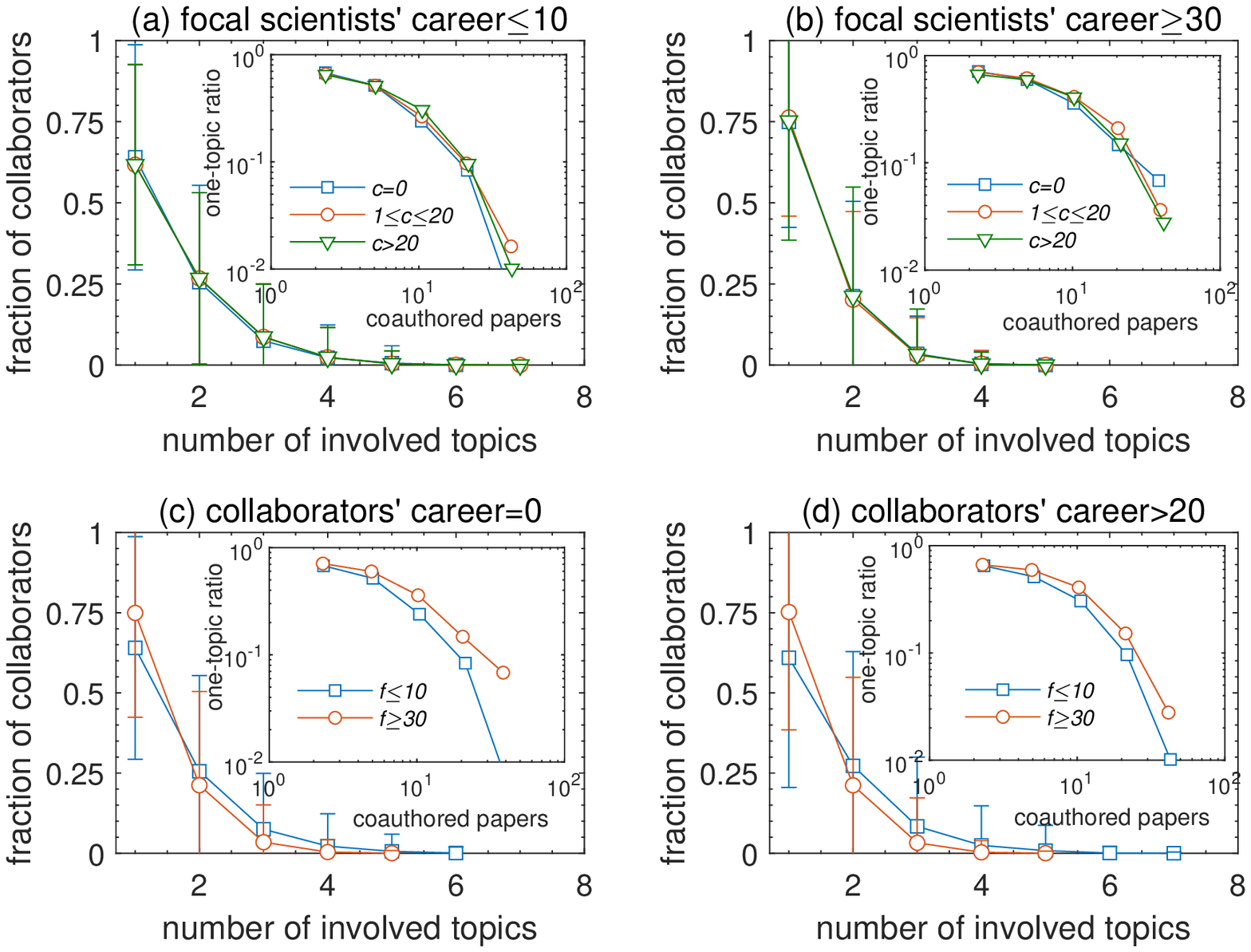}\\
   \textbf{Figure S11. Collaborator career vs focal scientist career.} (a) The distribution of the number of topics that collaborators are involved in. Inset: the fraction of single-topic collaborators who coauthored different number of papers with the focal scientist. We take only the collaborators who start collaborating with the focal scientist in the early career of the focal scientist (i.e. focal scientist's career$\leq$ 10 years). The different curves in the figures represent results of collaborators at different career stages when they start collaboration with the focal scientist. One can see that the results are consistent when selecting collaborators of different career lengths. (b) The same as panel (a), but we take only the collaborators who start collaborating with the focal scientist in the later career of the focal scientist (i.e. focal scientist's career$\geq$ 30 years). Again, collaborators at different career stages show similar distributions, suggesting the tendency of single-topic collaboration in scientific research. (c) The same as panel (a), but we take only the collaborators who have no publication before they start collaboration with the focal scientist. We compare these collaborators at different career stages of the focal scientists, finding that collaborators who start collaboration in later career of the focal scientist tend to involve in fewer topics. (d) The same as panel (c), but we take only the collaborators who have already at least 20 years career before they start collaboration with the focal scientist. We find that the results are consistent with panel (c). 
\end{figure}

\clearpage
\begin{figure}[h!]
  \centering
  \includegraphics[width=\textwidth]{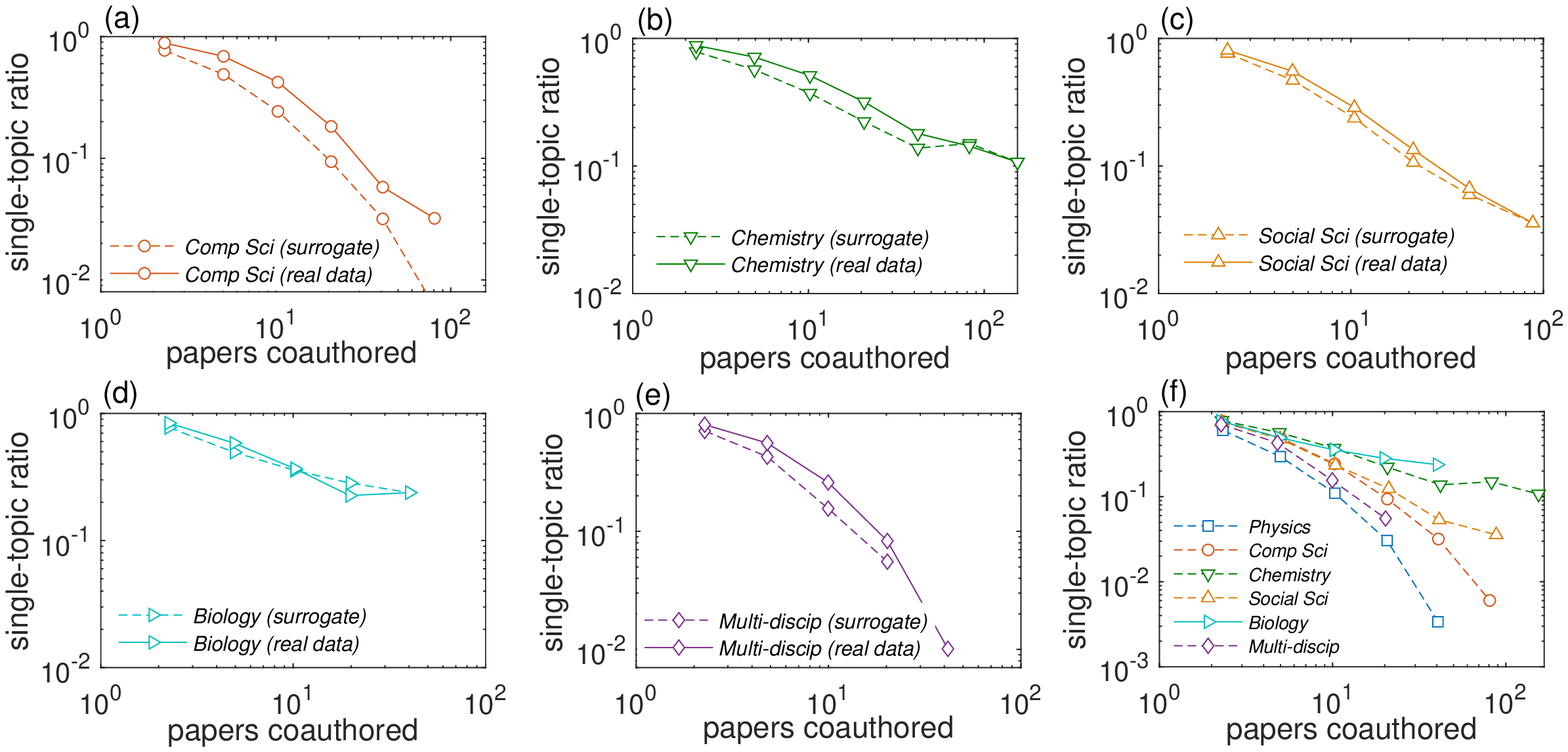}\\
   \textbf{Figure S12. The surrogate control in other disciplines.} The controlled surrogate is done by randomly shuffling the relations between collaborators and their coauthored papers with the focal scientist. Only the papers published in the same year are allowed to be shuffled in the randomization. In panel (a)-(e), we compare the real data with the surrogate control in the fraction of single-topic collaborators for the collaborators who coauthored different number of papers with the focal scientist. The disciplines are (a) Computer Science, (b) Chemistry, (c) Social Science, (d) Biology, (e) Multi-disciplinary Science. In panel (f), we compare the results of different disciplines in the surrogate control. We find that the fraction of single-topic collaborators in biology and chemistry is the highest when testing the surrogate control in these disciplines.
\end{figure}

\clearpage
\begin{figure}[h!]
  \centering
  \includegraphics[width=\textwidth]{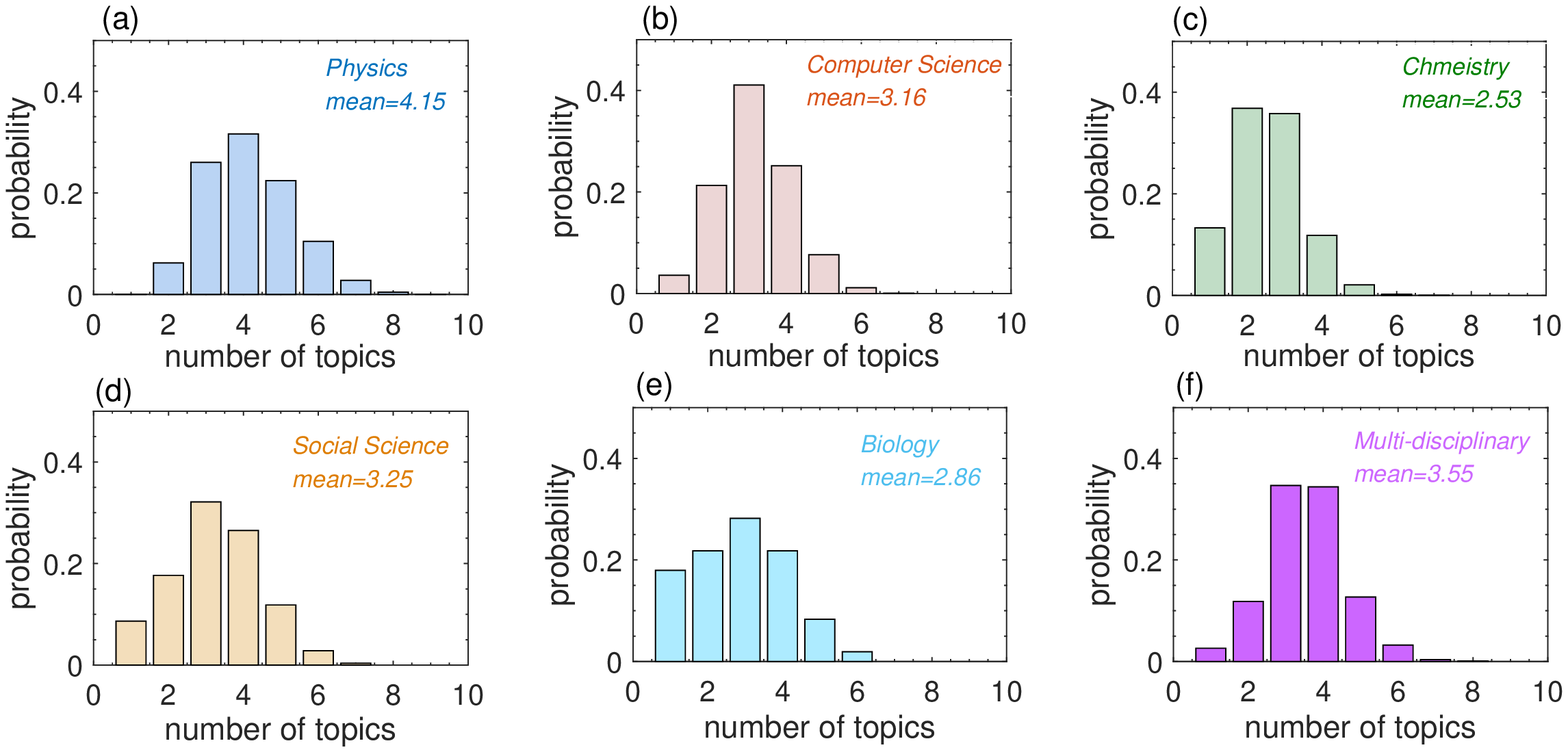}\\
   \textbf{Figure S13. Number of topics of scientists from different disciplines.} For each scientist in a discipline, we construct a co-citing network in which each node is a paper authored by this scientist and two papers have a link if they share at least one reference. The communities in the co-citing networks are detected via the fast-unfolding algorithm, with each significant community (more than 5\% papers) representing a major topic of the scientist. As the co-citing network needs to be large enough to ensure meaningful community detection results, we consider in each discipline data set only the scientists with at least 50 papers. We show the distribution of scientists' topic number in each discipline, (a) Physics, (b) Computer Science, (c) Chemistry, (d) Social Science, (e) Biology, (f) Multi-disciplinary Science. The mean of the topic number distribution is given in each panel. The results show that scientists from Chemistry and Biology focus on fewer topics than the scientists from other considered disciplines.
\end{figure}

\clearpage
\begin{figure}[h!]
  \centering
  \includegraphics[width=\textwidth]{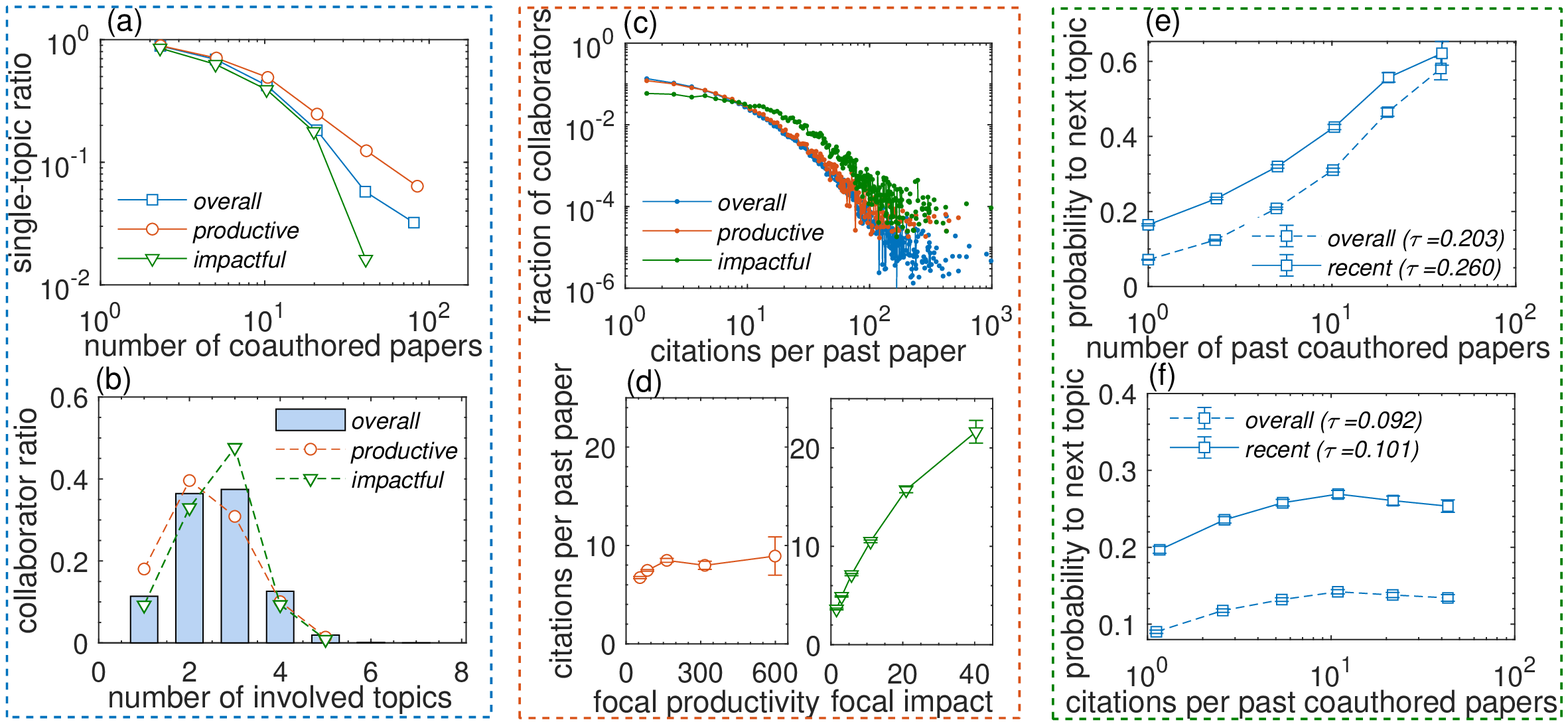}\\
   \textbf{Figure S14. Results of computer science data.} (a) The fraction of single-topic collaborators for the collaborators who coauthored different number of papers with the focal scientist. We compare the 5\% most productive scientists and the 5\% most impactful scientists. (b) The distribution of the number of topics for the collaborators who coauthored at least 10 papers with the focal scientists. The productive scientists have higher fraction of single-topic collaborators while the highly cited scientists have lower fraction of single-topic collaborators. (c) The distribution of citations per past paper of collaborators when they started collaboration with the focal scientist. The distribution of impactful scientists exhibits a fatter tail than the overall case, suggesting that impactful scientists are associated with impactful initial collaborators. (d) The average citations per past paper of collaborators for focal scientists with different productivity or impact. The average citations per past paper of collaborators have an significant increasing trend with the impact of the focal scientists. (e) The probability of existing collaborators to be involved in the new topic of the focal scientist as a function of the number of past coauthored papers. We compute also the probability among recent collaborators who have coauthored papers with the focal scientists within the past two years. The mean probability of the overall case is 0.111 and the mean probability of the recent case is 0.237. The legend shows also the kendall's $\tau$ correlation between the probability to join next topic and the number of past coauthored papers. (f) The probability of a collaborator to join the next topic of the focal scientist versus the mean citations of their past coauthored papers. Both the overall probability and the probability among recent collaborators show an increasing trend. The legend shows also the kendall's $\tau$ correlation between the probability to join next topic and the mean citations of past coauthored papers.
\end{figure}

\clearpage
\begin{figure}[h!]
  \centering
  \includegraphics[width=\textwidth]{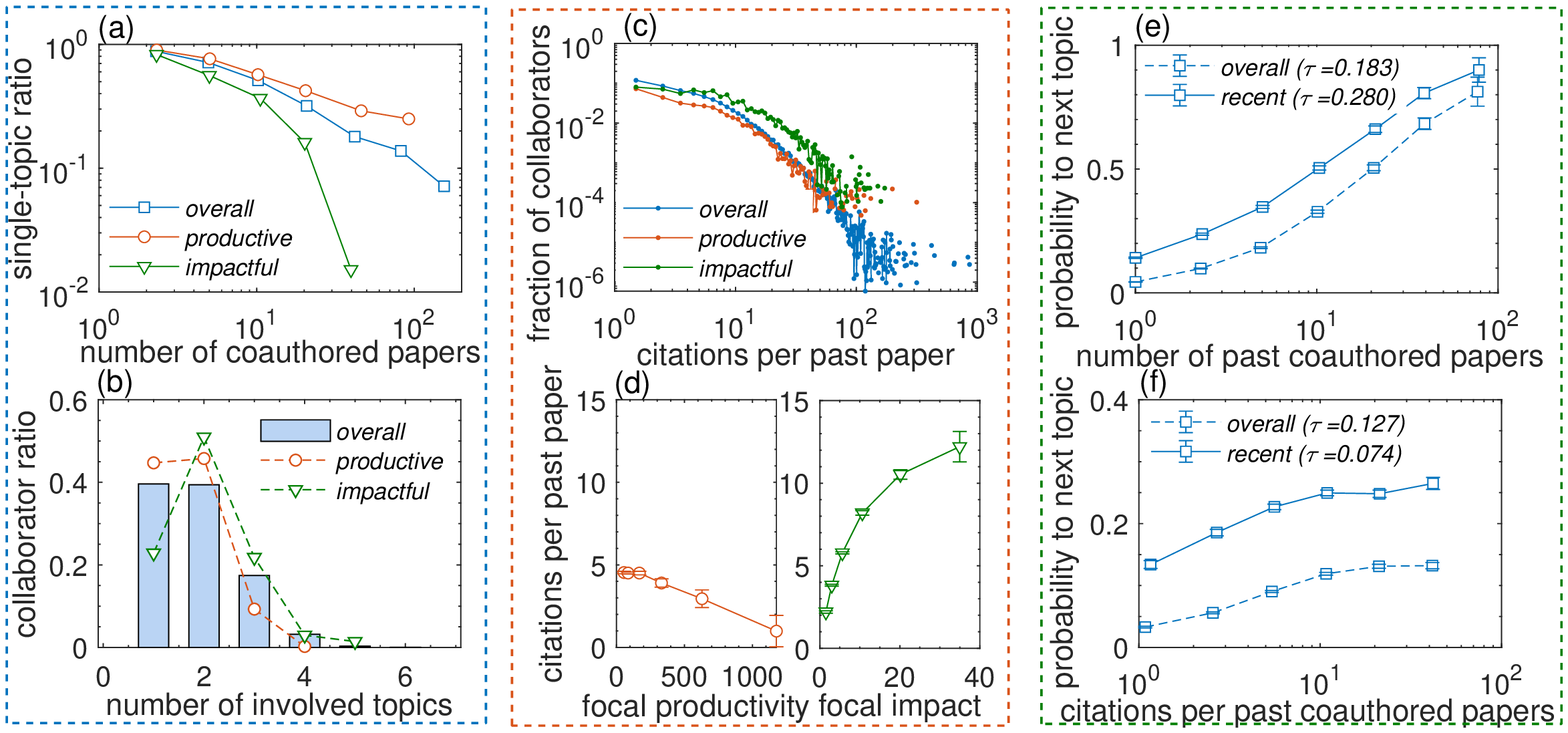}\\
   \textbf{Figure S15. Results of chemistry data.} (a) The fraction of single-topic collaborators for the collaborators who coauthored different number of papers with the focal scientist. We compare the 1\% most productive scientists and the 1\% most impactful scientists. (b) The distribution of the number of topics for the collaborators who coauthored at least 10 papers with the focal scientists. The productive scientists have higher fraction of single-topic collaborators while the highly cited scientists have lower fraction of single-topic collaborators. (c) The distribution of citations per past paper of collaborators when they started collaboration with the focal scientist. The distribution of impactful scientists exhibits a fatter tail than the overall case, suggesting that impactful scientists are associated with impactful initial collaborators. (d) The average citations per past paper of collaborators for focal scientists with different productivity or impact. The average citations per past paper of collaborators have an significant increasing trend with the impact of the focal scientists. (e) The probability of existing collaborators to be involved in the new topic of the focal scientist as a function of the number of past coauthored papers. We compute also the probability among recent collaborators who have coauthored papers with the focal scientists within the past two years. The mean probability of the overall case is 0.073 and the mean probability of the recent case is 0.219. The legend shows also the kendall's $\tau$ correlation between the probability to join next topic and the number of past coauthored papers. (f) The probability of a collaborator to join the next topic of the focal scientist versus the mean citations of their past coauthored papers. Both the overall probability and the probability among recent collaborators show an increasing trend. The legend shows also the kendall's $\tau$ correlation between the probability to join next topic and the mean citations of past coauthored papers.
\end{figure}

\clearpage
\begin{figure}[h!]
  \centering
  \includegraphics[width=\textwidth]{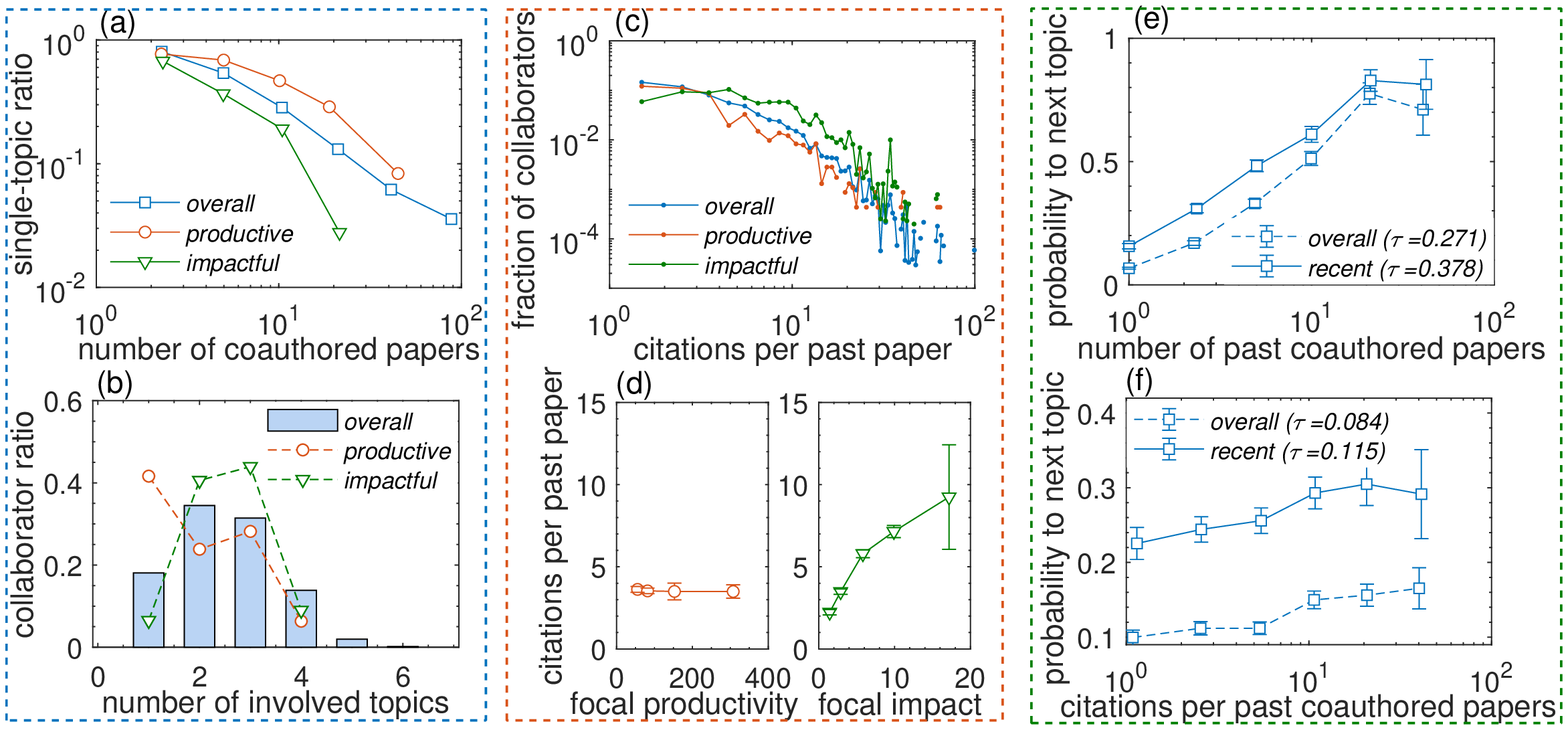}\\
   \textbf{Figure S16. Results of social science data.} (a) The fraction of single-topic collaborators for the collaborators who coauthored different number of papers with the focal scientist. We compare the 5\% most productive scientists and the 5\% most impactful scientists. (b) The distribution of the number of topics for the collaborators who coauthored at least 10 papers with the focal scientists. The productive scientists have higher fraction of single-topic collaborators while the highly cited scientists have lower fraction of single-topic collaborators. (c) The distribution of citations per past paper of collaborators when they started collaboration with the focal scientist. The distribution of impactful scientists exhibits a fatter tail than the overall case, suggesting that impactful scientists are associated with impactful initial collaborators. (d) The average citations per past paper of collaborators for focal scientists with different productivity or impact. The average citations per past paper of collaborators have an significant increasing trend with the impact of the focal scientists. (e) The probability of existing collaborators to be involved in the new topic of the focal scientist as a function of the number of past coauthored papers. We compute also the probability among recent collaborators who have coauthored papers with the focal scientists within the past two years. The mean probability of the overall case is 0.118 and the mean probability of the recent case is 0.255. The legend shows also the kendall's $\tau$ correlation between the probability to join next topic and the number of past coauthored papers. (f) The probability of a collaborator to join the next topic of the focal scientist versus the mean citations of their past coauthored papers. Both the overall probability and the probability among recent collaborators show an increasing trend. The legend shows also the kendall's $\tau$ correlation between the probability to join next topic and the mean citations of past coauthored papers.
\end{figure}

\clearpage
\begin{figure}[h!]
  \centering
  \includegraphics[width=\textwidth]{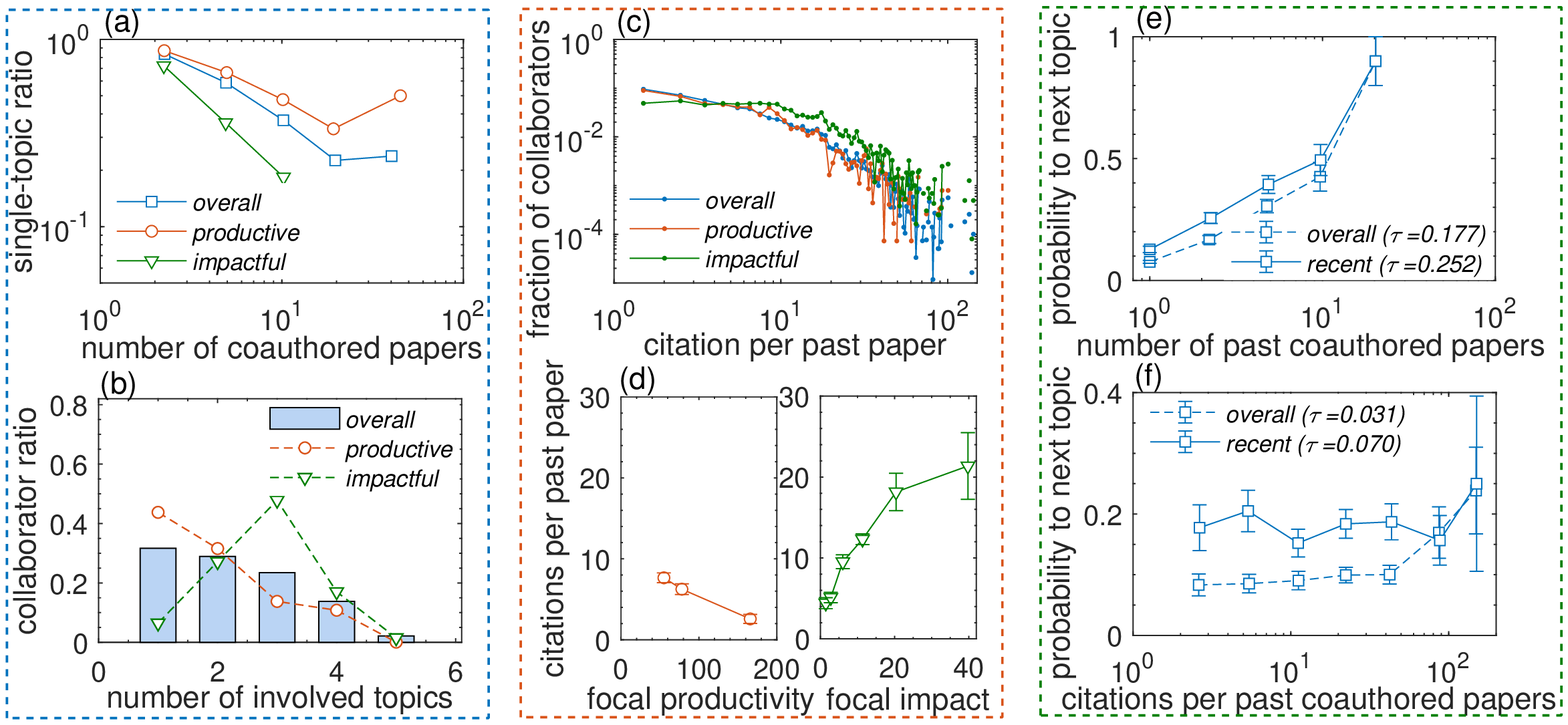}\\
   \textbf{Figure S17. Results of biology data.} (a) The fraction of single-topic collaborators for the collaborators who coauthored different number of papers with the focal scientist. We compare the 20\% most productive scientists and the 20\% most impactful scientists. (b) The distribution of the number of topics for the collaborators who coauthored at least 10 papers with the focal scientists. The productive scientists have higher fraction of single-topic collaborators while the highly cited scientists have lower fraction of single-topic collaborators. (c) The distribution of citations per past paper of collaborators when they started collaboration with the focal scientist. The distribution of impactful scientists exhibits a fatter tail than the overall case, suggesting that impactful scientists are associated with impactful initial collaborators. (d) The average citations per past paper of collaborators for focal scientists with different productivity or impact. The average citations per past paper of collaborators have an significant increasing trend with the impact of the focal scientists. (e) The probability of existing collaborators to be involved in the new topic of the focal scientist as a function of the number of past coauthored papers. We compute also the probability among recent collaborators who have coauthored papers with the focal scientists within the past two years. The mean probability of the overall case is 0.098 and the mean probability of the recent case is 0.177. The legend shows also the kendall's $\tau$ correlation between the probability to join next topic and the number of past coauthored papers. (f) The probability of a collaborator to join the next topic of the focal scientist versus the mean citations of their past coauthored papers. Both the overall probability and the probability among recent collaborators show an increasing trend. The legend shows also the kendall's $\tau$ correlation between the probability to join next topic and the mean citations of past coauthored papers.
\end{figure}

\clearpage
\begin{figure}[h!]
  \centering
  \includegraphics[width=\textwidth]{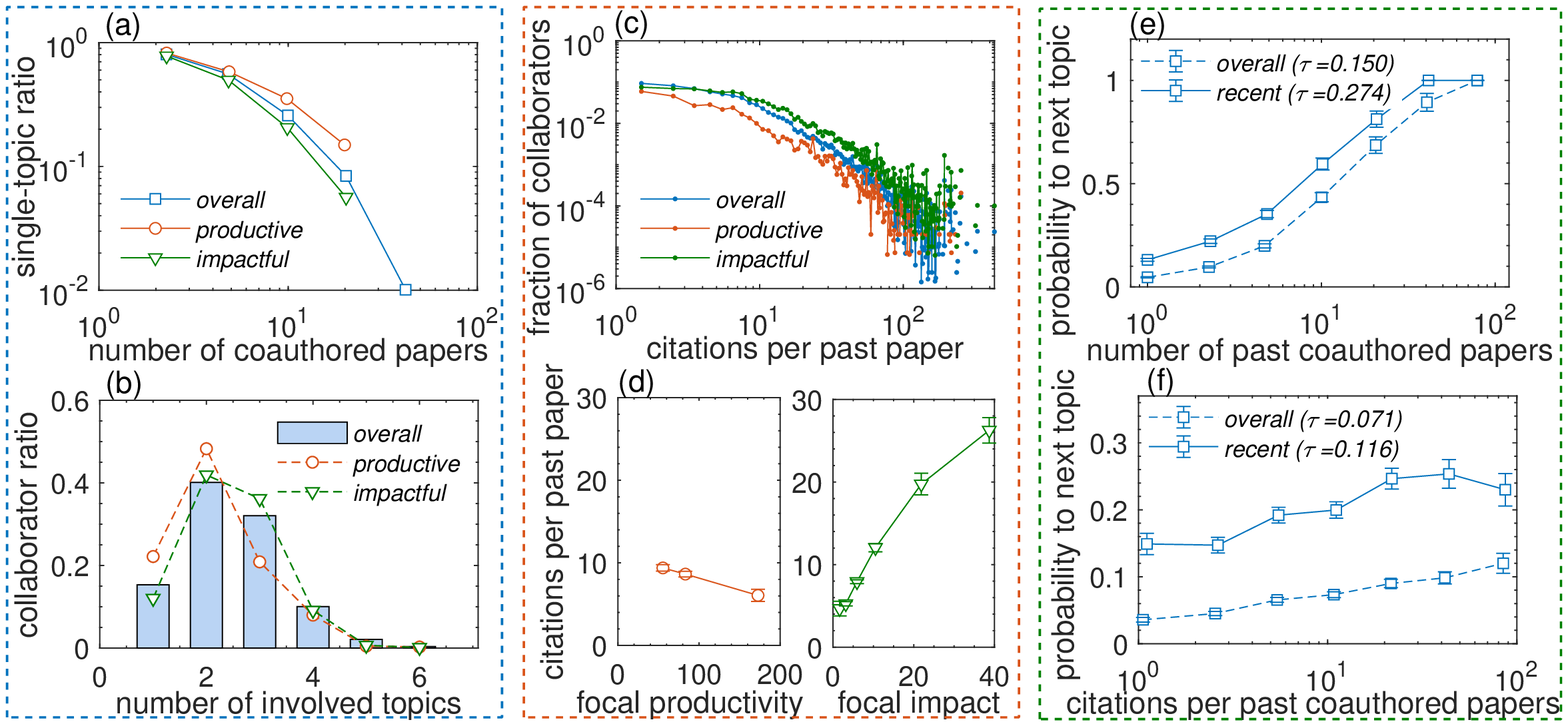}\\
   \textbf{Figure S18. Results of multidisciplinary science data.} (a) The fraction of single-topic collaborators for the collaborators who coauthored different number of papers with the focal scientist. We compare the 20\% most productive scientists and the 20\% most impactful scientists. (b) The distribution of the number of topics for the collaborators who coauthored at least 10 papers with the focal scientists. The productive scientists have higher fraction of single-topic collaborators while the highly cited scientists have lower fraction of single-topic collaborators. (c) The distribution of citations per past paper of collaborators when they started collaboration with the focal scientist. The distribution of impactful scientists exhibits a fatter tail than the overall case, suggesting that impactful scientists are associated with impactful initial collaborators. (d) The average citations per past paper of collaborators for focal scientists with different productivity or impact. The average citations per past paper of collaborators have an significant increasing trend with the impact of the focal scientists. (e) The probability of existing collaborators to be involved in the new topic of the focal scientist as a function of the number of past coauthored papers. We compute also the probability among recent collaborators who have coauthored papers with the focal scientists within the past two years. The mean probability of the overall case is 0.063 and the mean probability of the recent case is 0.173. The legend shows also the kendall's $\tau$ correlation between the probability to join next topic and the number of past coauthored papers. (f) The probability of a collaborator to join the next topic of the focal scientist versus the mean citations of their past coauthored papers. Both the overall probability and the probability among recent collaborators show an increasing trend. The legend shows also the kendall's $\tau$ correlation between the probability to join next topic and the mean citations of past coauthored papers.
\end{figure}

\clearpage
\section*{Supplementary Table}
\begin{table}[h!]
 \textbf{Table S1.} Kolmogorov-Smirnov test of the distributions of the features of collaborators before they join the first topic with a focal scientist (significant test for the results in Fig. S8 in the paper). The left part of the table is the test of the career years of collaborators. The middle part of the table is the test of the publications number of collaborators. The right part of the table is the test of the average citation per paper of collaborators. For each focal scientist, we calculate respectively the average career years, average publication number, average citation per paper of his/her collaborators before they started collaboration. The $p$-value shown in the table is obtained by performing the K-S test of the distributions of each of these metrics between different groups of scientists (i.e. overall, productive and impactful). The results indicate that the average citation per paper of impactful scientists' collaborators are significantly different from the overall and productive scientists' collaborators ($p<0.001$). However, one cannot see such differences when testing career years and publications ($p>0.1$).

\resizebox{\textwidth}{!}{%
\begin{tabular}{cccccccccccc}
\hline
\multicolumn{4}{c}{$p$-value of K-S test on career years} & \multicolumn{4}{|c}{$p$-value of K-S test on publications} & \multicolumn{4}{|c}{$p$-value of K-S test on citations}\\
\hline
\multicolumn{1}{c|}{group}      & overall & productive & impact  & \multicolumn{1}{|c|}{group}       & overall & productive & impact     & \multicolumn{1}{|c|}{group}       & overall & productive & impact \\
\hline
\multicolumn{1}{c|}{overall} & -         & 0.9980    & 0.6224        & \multicolumn{1}{|c|}{overall}  & -          & 0.8771     & 0.8750  & \multicolumn{1}{|c|}{overall}  & -          &0.0663     & 2.76$*10^{-8}$ \\
\multicolumn{1}{c|}{productive} &           & -         & 0.4239        & \multicolumn{1}{|c|}{productive}  &            & -          &0.8445  & \multicolumn{1}{|c|}{productive}  &           & -         & 5.55$*10^{-14}$ \\
\multicolumn{1}{c|}{impact} &           &           & -             & \multicolumn{1}{|c|}{impact}  &            &            & -     & \multicolumn{1}{|c|}{impact}  &            &            & -          \\
\hline
\end{tabular}%
}
\end{table}


\clearpage
\begin{thebibliography}{99}
\bibitem{the2007wuchty} S. Wuchty, B. F. Jones, B. Uzzi, The increasing dominance of teams in production of knowledge, \emph{Science} \textbf{316}, 1036-1039 (2007).
\bibitem{principles2014milojevic} S. Milojevic, Principles of scientific research team formation and evolution, \emph{Proc. Natl. Acad. Sci. USA} \textbf{111}, 3984-3989 (2014).
\bibitem{team2005guimera} R. Guimera, B. Uzzi, J. Spiro, L. Amaral, Team assembly mechanisms determine collaboration network structure and team performance, \emph{Science} \textbf{308}, 697-702 (2005).
\bibitem{large2019wu} L. Wu, D. Wang, J. A. Evans, Large teams develop and small teams disrupt science and technology, \emph{Nature} \textbf{566}, 378-382 (2019).
\bibitem{fresh2021zeng} A. Zeng, Y. Fan, Z. Di, Y. Wang, S. Havlin, Fresh teams are associated with original and multidisciplinary research,\emph{Nat. Hum. Behav.} \textbf{5}, 1314-1322 (2021).
\bibitem{science2018fortunato} S. Fortunato, et al. Science of science, \emph{Science} \textbf{359}, eaao0185 (2018).
\bibitem{the2017zeng} A. Zeng, et al. The science of science: from the perspective of complex systems, \emph{Phys. Rep.} \textbf{714-715}, 1-73 (2017).
\bibitem{the2001newman} M. E. J. Newman, The structure of scientific collaboration networks, \emph{Proc. Natl. Acad. Sci. USA} \textbf{98}, 404-409 (2001).
\bibitem{assortative2002newman} M. E. Newman, J. Assortative mixing in networks, \emph{Phys. Rev. Lett.} \textbf{89}, 208701 (2002).
\bibitem{motifs2011krumov} L. Krumov, et al. Motifs in co-authorship networks and their relation to the impact of scientific publications, \emph{Eur. phys. J. B} \textbf{84}, 535-540 (2011).
\bibitem{community2002grivan} M. Girvan, M. E. J. Newman, Community structure in social and biological networks, \emph{Proc. Natl. Acad. Sci. USA} \textbf{99}, 7821 (2002).
\bibitem{quantifying2015petersen} A. M. Petersen, Quantifying the impact of weak, strong, and super ties in scientific careers, \emph{Proc. Natl. Acad. Sci. USA} \textbf{112}, E4671-E4680 (2015).
\bibitem{collective2014shen} H.-W. Shen, A.-L. Barabasi, Collective credit allocation in science, \emph{Proc. Natl. Acad. Sci. USA} \textbf{111}, 12325-12330 (2014).
\bibitem{team2005guimera} R. Guimera, B. Uzzi, J. Spiro, L. Amaral, Team assembly mechanisms determine collaboration network structure and team performance, \emph{Science} \textbf{308}, 697-702 (2005).
\bibitem{understand2016klug} M. Klug, J. P. Bagrow, Understanding the group dynamics and success of teams. \emph{R. Soc. Open Sci.} \textbf{3}, 160007 (2016).
\bibitem{multinational2015hsiehchen} D. Hsiehchen, M. Espinoza, A. Hsieh, Multinational teams and diseconomies of scale in collaborative research, \emph{Sci. Adv.} \textbf{1}, e1500211 (2015).
\bibitem{evolution2016coccia} M. Coccia, L. Wang, Evolution and convergence of the patterns of international scientific collaboration, \emph{Proc. Natl. Acad. Sci. USA} \textbf{113}, 2057-2061 (2016).
\bibitem{multi2008jones} B. F. Jones, S. Wuchty, B. Uzzi, Multi-university research teams: Shifting impact, geography, and stratification in science, \emph{Science} \textbf{322}, 1259-1262 (2008).
\bibitem{the2010malmgren} R. D. Malmgren, J. M. Ottino, L. A. N. Amaral, The role of mentorship in protege performance. \emph{Nature} \textbf{465}, 622-626 (2010).
\bibitem{scientific2021jin} C. Jin, Y. Ma, B. Uzzi, Scientific prizes and the extraordinary growth of scientific topics, \emph{Nature Communications} \textbf{12}, 5619 (2021).
\bibitem{quantifying2017jia} T. Jia, D. Wang, B. K. Szymanski, Quantifying patterns of research-interest evolution, \emph{Nat. Hum. Behav.} \textbf{1}, 0078 (2017).
\bibitem{taking2019battiston} F. Battiston, et al. Taking census of physics, \emph{Nat. Rev. Phys.} \textbf{1}, 89-97 (2019).
\bibitem{increase2019zeng} A. Zeng, et al. Increasing trend of scientists to switch between topics, \emph{Nat. Commun.} \textbf{10}, 3439 (2019).
\bibitem{scientific2020li} J. Li, Y. Yin, S. Fortunato, D. Wang, Scientific elite revisited: patterns of productivity, collaboration, authorship and impact, \emph{J. R. Soc. Interface} \textbf{17}, 20200135 (2020).
\bibitem{understand2021liu} L. Liu, N. Dehmamy, J. Chown, C. L. Giles, D. Wang, Understanding the onset of hot streaks across artistic, cultural, and scientific careers, \emph{Nat. Commun.} \textbf{12}, 5392 (2021).
\bibitem{the2009jones} B. F. Jones, The burden of knowledge and the death of the Renaissance man: Is innovation getting harder? \emph{Rev. Econ. Stud.} \textbf{76}, 283-317 (2009).
\bibitem{age2011jones} B. F. Jones, \& Weinberg, B. A. Age dynamics in scientific creativity, \emph{Proc. Natl. Acad. Sci. USA} \textbf{108}, 18910-18914 (2011).
\bibitem{tradition2015foster} J. G. Foster, A. Rzhetsky, J. A. Evans, Tradition and innovation in scientists' research strategies, \emph{Am. Sociol. Rev.} \textbf{80}, 875-908 (2015).
\bibitem{atypical2013uzzi} B. Uzzi, S. Mukherjee, M. Stringer, B. Jones, Atypical combinations and scientific impact, \emph{Science} \textbf{342}, 468-472 (2013).
\bibitem{quantifying2016sinatra} R. Sinatra, et al. Quantifying the evolution of individual scientific impact, \emph{Science} \textbf{354}, aaf5239 (2016).
\bibitem{hot2018liu} L. Liu, et al. Hot streaks in artistic, cultural, and scientific careers, \emph{Nature} \textbf{559}, 396-399 (2018).
\bibitem{extraction2008tang} J. Tang, et al. ArnetMiner: Extraction and Mining of Academic Social Networks. In Proceedings of the Fourteenth ACM SIGKDD International Conference on Knowledge Discovery and Data Mining (SIGKDD'2008) 990-998 (2008).
\bibitem{an2015sinha} A. Sinha, et al. An Overview of Microsoft Academic Service (MAS) and Applications. In Proceedings of the 24th International Conference on World Wide Web (WWW 15 Companion). ACM, New York, NY, USA, 243-246 (2015).
\bibitem{fast2008blondel} V. D. Blondel, J.-L. Guillaume, R. Lambiotte and E. Lefebvre, Fast unfolding of communities in large networks, \emph{J. Stat. Mech.} P10008 (2008).
\bibitem{fast2004newman} M. E. J. Newman, Fast algorithm for detecting community structure in networks, \emph{Phys. Rev. E} 69, 066133 (2004).
\bibitem{adapt2022hill} R. Hill, Y. Yin, C. Stein, D. Wang, B. F. Jones, Adaptability and the Pivot Penalty in Science, arXiv:2107.06476 (2022)
\bibitem{reputation2013petersen} A. M. Petersen, et al., Reputation and impact in academic careers, \emph{PNAS} \textbf{111}, 15316-15321 (2013).
\end{thebibliography}
\end{document}